# Identification and properties of intense star-forming galaxies at redshifts *z>10*

*[1]Robertson, B. E., *[2,3]Tacchella, S., [4]Johnson, B. D., [5]Hainline, K., [5]Whitler, L., [4]Eisenstein D. J., [6]Endsley, R., [5]Rieke, M., [5]Stark, D. P., [5]Alberts, S., [7]Dressler, A., [5]Egami, E., [8]Hausen, R., [5]Rieke, G., [5]Shivaei, I., [9]Williams, C. C., [5]Willmer, C. N. A., [10]Arribas, S., [11,12]Bonaventura, N., [13]Bunker, A., [13]Cameron, A. J., [14]Carniani, S., [15]Charlot, S., [13]Chevallard, J., [2,3]Curti, M., [16]Curtis-Lake, E., [2,3]D'Eugenio, F., [11,12]Jakobsen, P., [2,3]Looser, T. J., [17]Lützgendorf, N., [2,3,18]Maiolino, R., [19]Maseda, M. V., [17]Rawle, T., [20]Rix, H.-W., [21]Smit, R., [2,3]Übler, H., [22]Willott, C., [2,3]Witstok, J., [23]Baum, S., [24]Bhatawdekar, R., [25,26]Boyett, K., [5]Chen, Z., [20]de Graaff, A., [5]Florian, M., [5]Helton, J. M., [5]Hviding, R. E., [5]Ji, Z., [27]Kumari, N., [5]Lyu, J., [28]Nelson, E., [2,3]Sandles, L., [13,18]Saxena, A., [1,29]Suess, K. A., [5]Sun, F., [5]Topping, M. & [13]Wallace, I. E. B.

[1]Department of Astronomy and Astrophysics, University of California, Santa Cruz, Santa Cruz, CA 95064, USA. [2]Kavli Institute for Cosmology, University of Cambridge, Madingley Road, Cambridge, CB3 0HA, UK. [3]Cavendish Laboratory, University of Cambridge, 19 JJ Thomson Avenue, Cambridge, CB3 0HE, UK. [4]Center for Astrophysics | Harvard & Smithsonian, 60 Garden Street, Cambridge, MA 02138, USA. [5]Steward Observatory, University of Arizona, 933 N. Cherry Avenue, Tucson, AZ 85721, USA. [6]Department of Astronomy, University of Texas, Austin, TX, 78712, USA. [7]The Observatories of the Carnegie Institution for Science, 813 Santa Barbara St., Pasadena, CA 91101, USA. [8]Department of Physics and Astronomy, The Johns Hopkins University, 3400 N. Charles St., Baltimore, MD 21218, USA. [9]NSF's National Optical-Infrared Astronomy Research Laboratory, 950 North Cherry Avenue, Tucson, AZ 85719, USA. [10]Centro de Astrobiología (CAB), CSIC–INTA, Cra. de Ajalvir Km.~4, 28850- Torrejón de Ardoz, Madrid, Spain. [11]Cosmic Dawn Center (DAWN), Copenhagen, Denmark. [12]Niels Bohr Institute, University of Copenhagen, Jagtvej 128, DK-2200, Copenhagen, Denmark. [13]Department of Physics, University of Oxford, Denys Wilkinson Building, Keble Road, Oxford OX1 3RH, UK. [14]Scuola Normale Superiore, Piazza dei Cavalieri 7, I-56126 Pisa, Italy. [15]Sorbonne Université, CNRS, UMR7095, Institut d'Astrophysique de Paris, 98 bis bd Arago, 75014 Paris, France. [16]Centre for Astrophysics Research, Department of Physics, Astronomy and Mathematics, University of Hertfordshire, Hatfield AL10 9AB, UK. [17]European Space Agency, Space Telescope Science Institute, Baltimore, Maryland, USA. [18]Department of Physics and Astronomy, University College London, Gower Street, London WC1E 6BT, UK. [19]Department of Astronomy, University of Wisconsin-Madison, 475 N. Charter St., Madison, WI 53706 USA. [20]Max-Planck-Institut für Astronomie, Königstuhl 17, D-69117, Heidelberg, Germany. [21]Astrophysics Research Institute, Liverpool John Moores University, 146 Brownlow Hill, Liverpool L3 5RF, UK. [22]NRC Herzberg, 5071 West Saanich Rd, Victoria, BC V9E 2E7, Canada. [23]Dept of Physics & Astronomy University of Manitoba, 30A Sifton Rd, Winnipeg MB R3T 2N2 Canada. [24]European Space Agency, ESA/ESTEC, Keplerlaan 1, 2201 AZ Noordwijk, NL. [25]School of Physics, University of Melbourne, Parkville 3010, VIC, Australia. [26]ARC Centre of Excellence for All Sky Astrophysics in 3 Dimensions (ASTRO 3D), Australia.[27]AURA for European Space Agency, Space Telescope Science Institute, 3700 San Martin Drive. Baltimore, MD, 21210. [28]Department for Astrophysical and Planetary Science, University of Colorado, Boulder, CO 80309, USA. [29]Kavli Institute for Particle Astrophysics and Cosmology and Department of Physics, Stanford University, Stanford, CA 94305, USA.
*These authors contributed equally to this work
Corresponding authors: Brant Robertson (brant@ucsc.edu), Sandro Tacchella (st578@cam.ac.uk)

**Surveys with James Webb Space Telescope (JWST) have discovered candidate galaxies in the first 400 Myr of cosmic time. Preliminary indications have suggested these candidate galaxies may be more massive and abundant than previously thought. However, without confirmed distances, their inferred properties remain uncertain. Here we identify four galaxies located in the JWST Advanced Deep Extragalactic Survey (JADES) Near-Infrared Camera (NIRCam) imaging with photometric redshifts $z \sim 10 - 13$. These galaxies include the first redshift $z > 12$ systems discovered with distances spectroscopically confirmed by JWST in a companion paper. Using stellar population modelling, we find the galaxies typically contain a hundred million solar masses in stars, in stellar populations that are less than one hundred million years old. The moderate star formation rates and compact sizes suggest elevated star formation rate surface densities, a key indicator of their formation pathways. Taken together, these measurements show that the first galaxies contributing to cosmic reionisation formed rapidly and with intense internal radiation fields.**

The properties of the earliest galaxies elucidate how the first cosmic structures form and evolve. Their nature constrains both the astrophysics associated with the baryonic components of galaxies and the cosmological physics that governs the development of their dark matter halos. Star-forming galaxies likely reionised intergalactic hydrogen during the first billion years of cosmic history, but the duration of the reionisation process and when it initiated are unknown[1]. The need to address these questions about the physics of galaxy formation and cosmic reionisation motivates the search for ever more distant galaxies. After the launch of JWST in December 2021, its near-infrared sensitivity and resolution could be applied to the search for and characterisation of high-redshift galaxies[2-19] starting with the first observations[20] in July 2022. JWST programmes scheduled early, including the Cosmic Evolution Early Release Science[21] (CEERS; JWST programme 1345) and the GLASS[22] (JWST programme 1324) surveys, yielded distant galaxy candidates putatively at redshifts $z \sim 12 - 17$ (refs. 23-28). Galaxy candidates with large claimed stellar masses ($M_\star \sim 10^{11} M_\odot$) and photometric redshifts as high as $z \sim 11$ have been identified[29] that could challenge standard models for galaxy formation[30-32]. However, none of these candidates have yet been confirmed spectroscopically. To date, the highest redshift spectroscopic observations with JWST extend only to redshift $z \sim 9.5$-$9.8$[19,33].

JADES, a collaborative program between the NIRCam and NIRSpec Guaranteed Time Observations (GTO) instrument science teams, was designed to both discover high-redshift galaxies with NIRCam[34] and spectroscopically confirm them with NIRSpec[35]. Initial JADES NIRCam observations of the Great Observatories Origins Deep Survey – South (GOODS-S; $\alpha = 53.165$ deg, $\delta = -27.786$ deg), including the Hubble Ultra Deep Field, were conducted starting September 29, 2022, under JWST programme 1180, acquiring NIRCam F090W, F115W, F150W, F200W, F277W, F335M, F356W, F410M, and F444W imaging ($\lambda \approx 0.8 - 5\ \mu m$), covering an area of 65 arcmin$^2$ to a depth of $f_{F200W} \approx 3.5 - 5$ nJy ($5\sigma$, 0.15" radius apertures; 30.0-29.7AB). A search for high-redshift galaxies was conducted by creating an inverse variance-weighted stacked detection image of the F200W through F444W filters and then selecting contiguous regions with flux higher than $3\sigma$/pix above the local mean background noise as objects. Forced photometry was performed at the locations of detected objects for all JADES filter images, public imaging from JWST programme 1963 in F182M, F210M, F430M, F460M, and F480M around the Ultra Deep Field, and publicly available Hubble Space Telescope (HST) Legacy Fields[36] (v2.0) F435W, F606W, F775W, F814W, F850LP, F105W, F125W, F140W, and F160W images ($\lambda \approx 0.4 - 1.8\ \mu m$). Photometric redshifts were estimated for every object in the catalogue to identify high-redshift galaxy candidates. A planned JADES spectroscopic campaign was then conducted starting October 20, 2022 with the JWST/NIRSpec microshutter array (MSA), opening slits at the positions of NIRCam-selected galaxy candidates that could be accommodated by pre-set pointings of the MSA. Other high-redshift galaxy candidates selected from prior HST imaging were included in the masks, in addition to a broader selection of known and candidate galaxies at lower redshifts. The reduction, analysis and interpretation of the spectroscopic data are presented in a companion paper (ref. 37). Confirmation of $z > 10$ candidates was possible for two HST-selected targets and two galaxies selected by our JWST/NIRCam imaging, with spectroscopic redshifts[37] from the Lyman-$\alpha$ break and other properties presented in Table 1 and the multiband photometry reported in Extended Data Table 1. The typical F200W fluxes of these distant galaxies are $f_{F200W} \approx 10$ nJy ($m \approx 29$ AB mag), and they are also well-detected in redder bands ($SNR \sim 5 - 20$). Given their distances, their ultraviolet magnitudes are between $M_{UV} \approx -18.4$ for the intrinsically faintest and $M_{UV} \approx -19.3$ for the brightest. These four objects are the focus of the work presented here.

Figure 1 presents JWST F200W-F150W-F115W colour images of the most distant spectroscopically confirmed galaxies in our sample, overlaid on a JWST F444W-F200W-F115W colour mosaic. The orange-red colours of these galaxies in the images reflect the near-complete absorption of light in the F115W and potentially F150W filters by intervening neutral hydrogen in the intergalactic medium. The two JWST/NIRCam-selected sources (JADES-GS-z13-0 and JADES-GS-z12-0), newly discovered here, are the most distant spectroscopically-confirmed galaxies known, with redshifts of $z = 13.20$ and $z = 12.63$ (ref. 37). Of the HST-discovered galaxies, our object JADES-GS+53.16476-27.77463 (JADES-GS-z11-0) is notably also known as UDFj-39546284. This galaxy was identified[38] as a $z\sim10$ galaxy from HST WFC3 imaging before receiving further observations[39] that constrained the photometric redshift to be $z \approx 11.9^{+0.3}_{-0.5}$. NIRCam medium band imaging in five filters have been used to estimate[40] the photometric redshift as $z \approx 12.0^{+0.1}_{-0.2}$. Here, the JWST data without our nine additional NIRCam filters provided a photometric redshift estimate of $z \approx 11.7^{+0.5}_{-0.4}$ and ref. 37 newly confirm that the galaxy lies at a spectroscopic redshift of $z = 11.58$. The final HST-selected object (JADES-GS-z10-0, originally known as UDFj-38116243; ref. 38) is now spectroscopically confirmed by ref. 37 to lie at redshift $z = 10.38$. The distribution of these sources across the JADES field arises from the placement of the NIRSpec/MSA footprint and does not reflect a possible physical clustering.

To accurately measure the photometric and morphological properties of these galaxies, we conduct forward modelling of the light distribution of each source and its neighbours on the sky using the *forcepho* code [BDJ in prep]. Figure 2 illustrates the photometric modelling process performed for each JWST filter simultaneously. At the location of sources detected in the stacked images, Sérsic[41] profiles with variable amplitude, axis ratio, position angle, and half-light radius are fit to pixel-level fluxes in the individual exposures (see Extended Data Figure 1 for detailed results). This approach avoids uncertainties in the effective PSF of the stack and pixel correlations that impact the uncertainties of aperture fluxes measured from the mosaics. The success of the fitting procedure can be judged from the residual images after subtracting the models from the data, and in each case, the model reliably captures the morphological complexity of the sources. Photometry measured from the forward-modelling procedure is consistent with the fluxes measured within 0.15" radii apertures, corrected to a total flux estimate using the JWST point-spread function (PSF) and the *forcepho* model. The image thumbnails reveal the compact sizes of these objects, with intrinsic half-light angular sizes $\theta_{1/2} \lesssim 0.015 - 0.04"$ as measured from forward-modelling of the light profiles and corresponding to physical scales of $r_{1/2} \lesssim 50 - 165$ pc at these redshifts. While the objects are small, their light profiles are all consistent with exponentials with semi-major to semi-minor axis ratios of $b/a \approx 0.6 - 0.9$.

Figure 2 also shows the results of Bayesian photometric modelling that allows for the inference of physical galaxy properties. The photometry and flux uncertainties are supplied to the *Prospector* code[42], and then the spectral energy distribution (SED) of each galaxy is fit assuming a flexible star formation history (SFH), a collection of stellar evolutionary models, and a fixed stellar initial mass function. The SED models allow for nebular continuum and line emission powered by Lyman continuum photons emitted by young stars. Extended Data Figure 2 shows the SED model fits while Figure 3 shows constraints on the physical properties of the galaxies, including the photometric redshift distributions from NIRCam imaging that are consistent with the galaxy spectroscopic redshifts. The marginalised distribution of the parameter constraints on the stellar mass, star formation rates, and stellar ages are also shown. In addition to the SFH, which reflects the stellar ages and integrates to give the stellar mass, the model parameters also include the stellar- and gas-phase metallicities, dust attenuation, and the interstellar medium ionisation parameter. The precise NIRSpec spectroscopic redshifts[37] significantly reduce the allowed parameter range given the photometric uncertainties. Inferred properties of the galaxies are listed in Table 1 while the results of the full likelihood analysis for each object are presented in Extended Data Figures 3-6.. Relative to the solar mass $M_\odot$, we find that the stellar masses of the objects are in the range $log_{10} M_\star / M_\odot \approx 7.8 - 8.9$ and the stellar ages, defined as the time before the object redshift at which half the stellar mass formed, are in the range $t_\star \approx 16 - 71$ Myr with $2\sigma$ upper limits of $t_\star \lesssim 167$ Myr. Their star formation rates are in the range $SFR \approx 0.2 - 5 \, M_\odot/yr$. These properties are consistent with expectations for galaxy formation in Λ-cold dark matter cosmologies at these redshifts[43,44]. These galaxies have median star formation rates comparable to the present Milky Way despite them being more than a hundred times less massive. These systems have sustained SFR more than ten times that of the Small Magellanic Cloud, a present-day galaxy of comparable stellar mass[45]. In the Methods, the stellar and gas-phase metallicities are shown to be poorly constrained but consistent with being under a tenth of the solar value, and the visual dust attenuation in these galaxies is $A_V \lesssim 0.3$ mag, with large uncertainty. The integrated yield of prior Type II supernovae is enough to enrich these galaxies with metals provided at least ≈10% of ejected metals remain and that the SFH has been sustained longer than 10 Myr[46]. Combining SFR with the size constraints, we find that the SFR surface densities of the galaxies are typically in the range $\Sigma_{SFR} \approx 15 - 180 \, M_\odot \, yr^{-1} kpc^{-2}$, which rival the most vigorous starbursts in the local universe[47]. Observations of Lyman continuum leakers at low redshift[48,49] and simple models of Lyman continuum escape[50] suggest that at such high SFR densities the escape fraction in our galaxy sample could be $f_{esc} \gtrsim 0.5$, consistent with our stellar population modelling, which may be sufficient for these galaxies to initiate the cosmic reionisation process.

With the rest-frame ultraviolet photometry enabled by JWST NIRCam and the first verification of the distances to these early galaxies from NIRSpec[37], the rest-frame ultraviolet luminosities of these objects can be computed and used to constrain the evolving cosmic SFH. Our sample was selected based on the available NIRSpec MSA coverage, and since it does not represent a complete sample of the highest-redshift galaxies we can only place lower limits on the total, volume-integrated star formation rate densities at z~13. Given the areal overlap between the MSA pointing and the NIRCam imaging and the photometric redshift selection, we expect the volume probed at $12.5 < z < 13.5$ to be $V \sim 14000 \, Mpc^3$. Using the two galaxies in this redshift range, we estimate a lower limit on the comoving cosmic SFR density from the star formation rate SFR$_{30}$ averaged over the preceding 30 Myr as $\rho_{SFR} \gtrsim 1.6 \times 10^{-4} \, M_\odot \, yr^{-1} Mpc^{-3}$. Using our stellar population modelling to estimate self-consistently the production rate of ionising photons and assuming $f_{esc} \sim 0.5$, we find $N_{ion} \gtrsim 3.3 \times 10^{49} s^{-1} Mpc^{-3}$. At these redshifts the intergalactic medium recombination timescale is $t_{rec} \approx 65 - 115 \, Myr$, and the observed galaxies could only manage to ionise their local intergalactic media[1]. These simple estimates indicate that the earliest forming galaxies identified in the

JADES imaging and verified through JADES spectroscopy are among the very first agents of cosmic reionisation, and will contribute to the eventual phase change from a neutral to ionised intergalactic medium at a later cosmic time.

The discovery and confirmation of galaxies at redshifts $z \sim 13$ provide a new foothold in the exploration of the distant universe. The stellar populations in these galaxies are very young, with ages comparable to the $t_{dyn} \sim 10 - 30\ Myr$ dynamical times within their star-forming regions. The stellar masses, star formation rates, and sizes imply that these galaxies may also be sources of Lyman continuum emission needed to initiate the cosmic reionisation process. Understanding the population of such galaxies, which will continue to be discovered and confirmed with JWST via JADES and other programmes, will require more extensive selection, completeness estimates to derive luminosity functions, and stellar population synthesis modelling to determine the distribution of stellar mass and star formation rate at the earliest times.

**Table 1 | Properties of Confirmed Redshift z>10 Galaxies**

| Object ID | Name | R.A. (deg) | Decl. (deg) | $z_{phot}$ | $z_{spec}$ | $m_{F200W}$ (AB mag) | $M_{UV}$ (AB mag) | $\beta_{UV}$ | $\log_{10} M_\star$ ($M_\odot$) | $\log_{10}$ SFR ($M_\odot/yr$) | $t_\star$ (Myr) |
|---|---|---|---|---|---|---|---|---|---|---|---|
| JADES-GS+53.14988-27.77650 | JADES-GS-z13-0 | 53.149880 | -27.776500 | $12.9^{+0.5}_{-0.4}$ | $13.20^{+0.04}_{-0.07}$ | 29.43±0.14 | -18.5±0.2 | $-2.44^{+0.15}_{-0.16}$ | $7.8^{+0.4}_{-0.5}$ | $0.0^{+0.3}_{-0.3}$ | $16^{+43}_{-13}$ |
| JADES-GS+53.16634-27.82156 | JADES-GS-z12-0 | 53.166338 | -27.821555 | $13.0^{+0.5}_{-0.5}$ | $12.63^{+0.24}_{-0.08}$ | 28.99±0.10 | -18.8±0.1 | $-2.18^{+0.13}_{-0.15}$ | $8.4^{+0.4}_{-0.7}$ | $0.1^{+0.4}_{-0.5}$ | $50^{+60}_{-44}$ |
| JADES-GS+53.16476-27.77463 | JADES-GS-z11-0 | 53.164763 | -27.774626 | $11.7^{+0.5}_{-0.4}$ | $11.58^{+0.05}_{-0.05}$ | 28.38±0.12 | -19.3±0.1 | $-2.06^{+0.12}_{-0.09}$ | $8.9^{+0.2}_{-0.4}$ | $0.3^{+0.4}_{-1.0}$ | $71^{+55}_{-45}$ |
| JADES-GS+53.15884-27.77349 | JADES-GS-z10-0 | 53.158836 | -27.773492 | $10.8^{+0.3}_{-0.3}$ | $10.38^{+0.07}_{-0.06}$ | 29.05±0.10 | -18.4±0.1 | $-2.42^{+0.09}_{-0.10}$ | $7.9^{+0.3}_{-0.5}$ | $0.0^{+0.2}_{-0.3}$ | $31^{+50}_{-21}$ |

Columns: (1) JADES Object identifier, (2) Object name, (3) Right Ascension (RA) in J2000, (4) Declination (Decl.) in J2000, (5) Photometric redshift ($z_{phot}$) from NIRCam photometry, (6) Spectroscopic redshift ($z_{spec}$) from ref. 37, (7) Apparent AB magnitude in the NIRCam F200W filter, (8) Absolute ultraviolet AB magnitude ($M_{UV}$), (9) Ultraviolet spectral slope ($\beta_{UV}$), (10) Stellar mass in $\log_{10}$ solar masses ($\log_{10} M_\star$), (11) Star formation rate in $\log_{10}$ solar masses per year ($\log_{10}$ SFR) averaged on 30 Myr timescales, (12) Half-mass stellar age in Myr ($t_\star$).

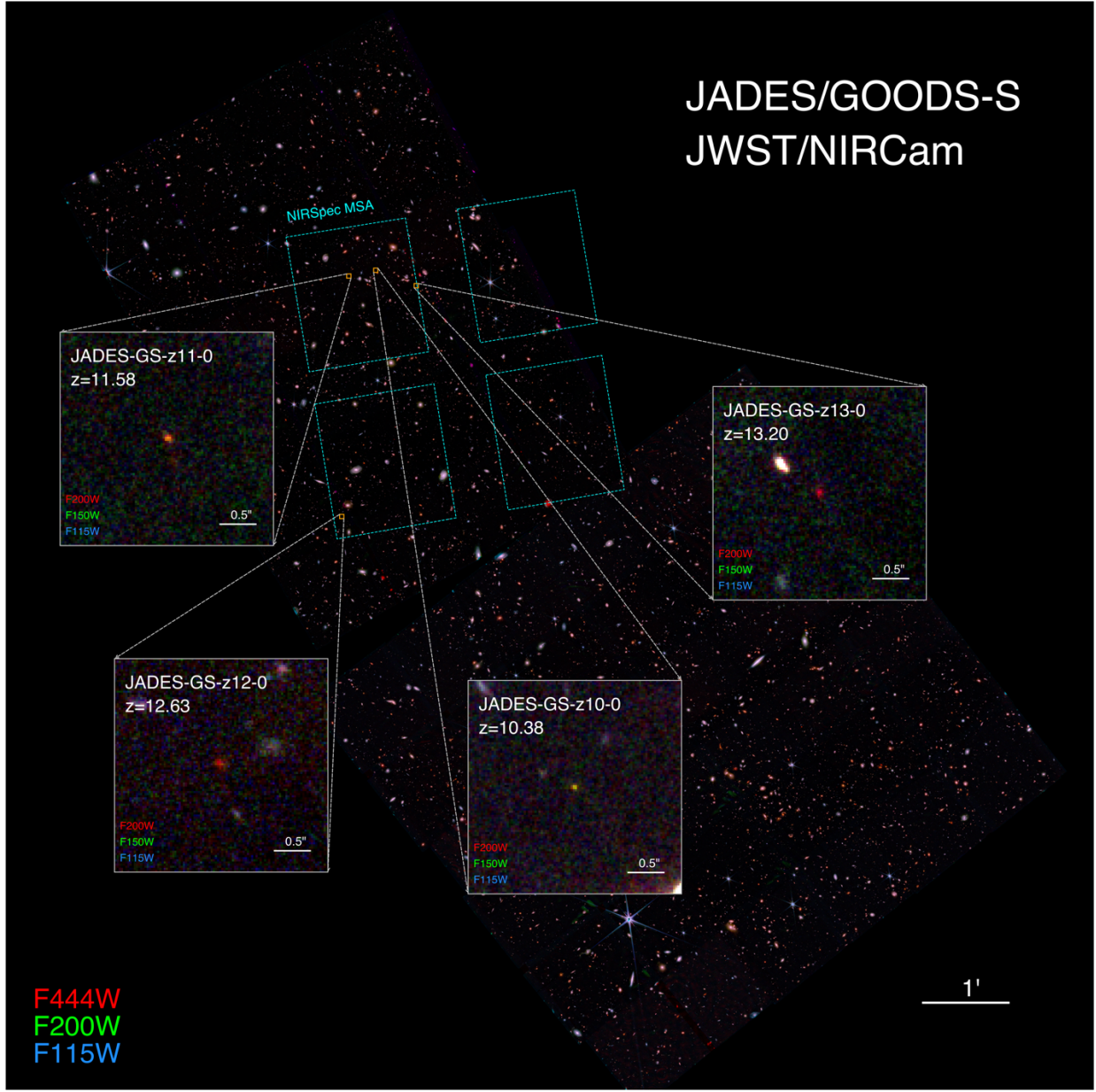

**Fig. 1| Distant galaxies selected and confirmed by the JWST JADES program.** From the JWST NIRCam imaging at wavelengths $\lambda \approx 0.8 - 5 \ \mu m$ (F444W-F200W-F115W shown as a colour mosaic, **a**), galaxies with photometrically-determined redshift estimates of $z_{phot} > 10$ were selected for JWST NIRSpec MSA follow-up (footprint in cyan). An initial sample of four $z>10$ galaxies (F200W-F150W-F115W thumbnails, **b-e**) was spectroscopically confirmed by ref. 37 at redshifts $z$~10.4-13.2. The most distant galaxies at $z$=13.20 and $z$=12.63 are newly discovered by JADES NIRCam imaging, while the $z$=10.38 and $z$=11.58 galaxies confirm previous photometric redshift estimates from the literature. The yellow-orange-red colours reflect the absorption of the F115W and F150W fluxes of these distant galaxies by the intervening intergalactic medium.

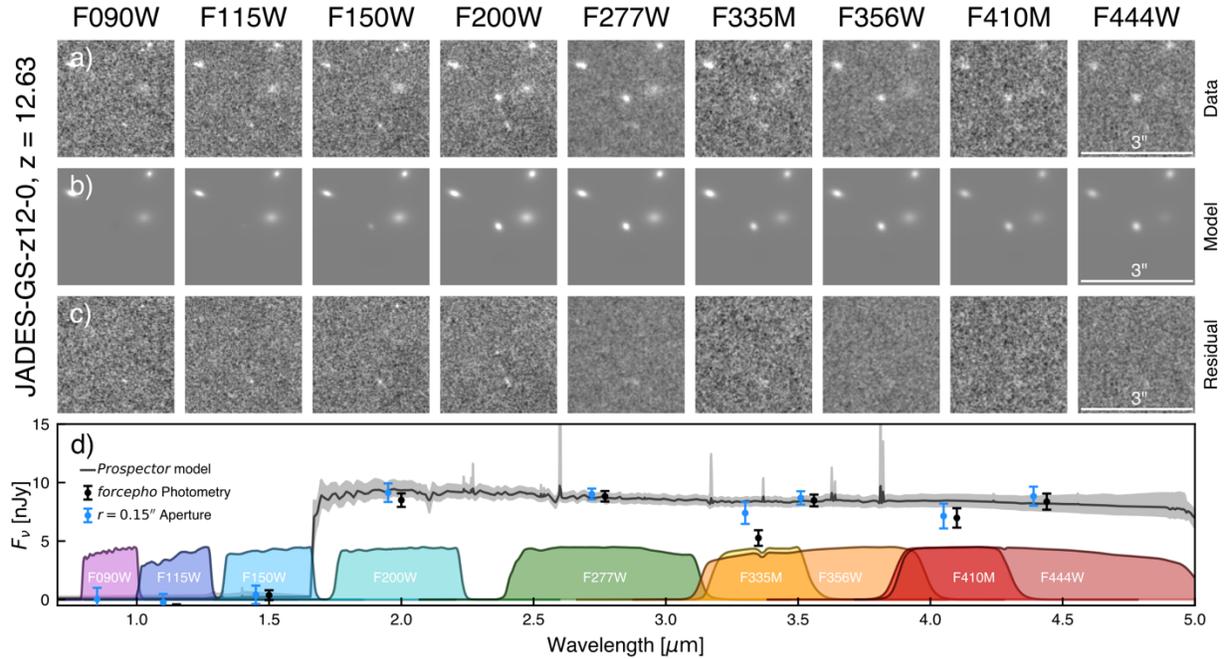

**Fig. 2| Precision photometry and spectral energy distribution (SED) modelling of JADES-GS-z12-0.** The sources detected by the JADES photometric pipeline are supplied to the *forcepho* scene modelling code, which then fits surface brightness profile models to the flux image of the object and its neighbours (**a**) in all NIRCam bands (columns, and filter curves in **d**) simultaneously. This method allows for the construction of accurate models of the observations (**b**) that leave only slight residuals (**c**) relative to the data. The source fluxes are then used as constraints for SED fitting (maximum likelihood fit to the $N=6$ detections, line in **d** with $1\sigma$ s.d. marginalised credibility interval as shaded region) to the photometry (points show mean values with $1\sigma$ s.d. error bars, including *forcepho* photometry in black and $r=0.15''$ aperture photometry in blue with an offset for clarity), which in turn constrain the physical galaxy properties. The postage stamps are 3" x 3", or about 13 kpc x 13 kpc at redshift $z=10$.

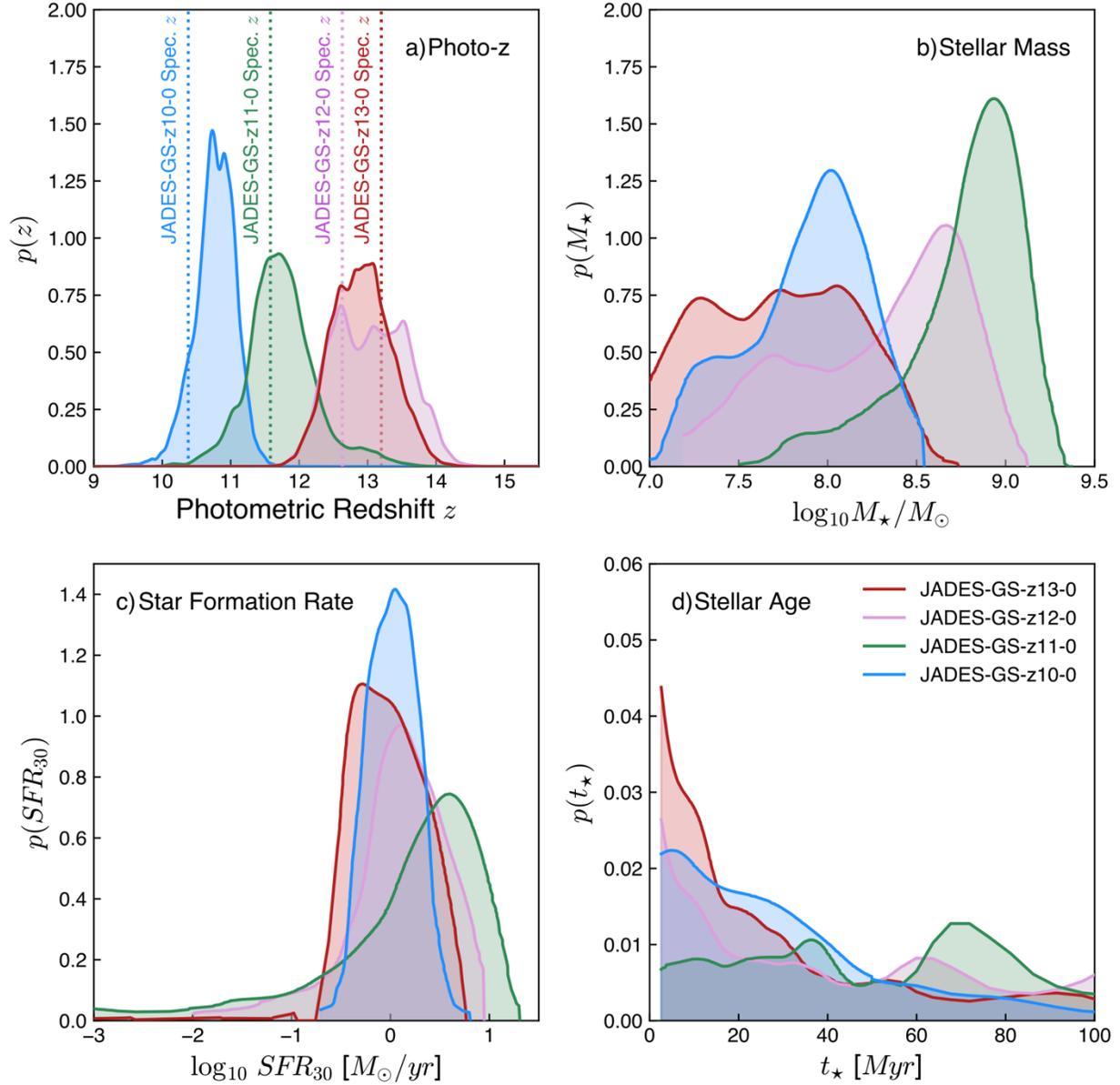

**Fig. 3 | Physical properties inferred from NIRCam imaging of distant, confirmed galaxies.** Multiband NIRCam imaging constrains the photometric redshift distributions of galaxies in our sample (**a**), which are consistent with the spectroscopic redshifts[37] (dashed lines). Using population synthesis models that feature a flexible star formation history, possible nebular continuum and line emission, stellar evolutionary modelling, and the effects of dust, the observed NIRCam photometry and NIRSpec spectroscopic redshifts are used to infer the stellar mass, star formation rate and history, and stellar age. Shown are the posterior distributions of the stellar mass (**b**), star formation rates (**c**), and mass-weighted median age of the stars (**d**) for the sample of z=10.4-13.2 galaxies presented here. We find a range of best-fit stellar masses of $log_{10}M_\star/M_\odot \sim 7.8 - 8.9$, star formation rates of $SFR \sim 0.2 - 5\ M_\odot/yr$, and ages of $t_\star \sim 16 - 71\ Myr$.

# Methods

**Survey Design and Motivation**
The observations reported here were carried out with the Near-Infrared Camera (NIRCam) on JWST, which was designed specifically for very deep multi-band surveying[34]. The camera features a field of view of 9.7 square arcmin for each of the short and long wavelength ranges, divided by a dichroic to allow surveying in both arms simultaneously. The Near-Infrared Spectrograph (NIRSpec) was also designed for this type of application, with a configurable multi-slit mask that allows spectra of many sources in a single exposure[35]. The JWST Advanced Deep Extragalactic Survey (JADES) is a collaboration of the NIRCam and NIRSpec teams to demonstrate and exploit these features by pooling over 750 hours of guaranteed time to conduct an ambitious study of galaxy evolution in the Great Observatories Origins Deep Survey (GOODS)-S and GOODS-N fields[51]. The JADES observing programme was designed to produce exquisite JWST imaging and spectroscopy of galaxy populations from cosmic noon (z ~ 2) to the redshift frontier (z>10), combining these data with the diverse deep multiwavelength imaging and spectroscopy already available in these two premier deep fields. JADES data will allow detailed study of many aspects of galaxy evolution such as the evolving demographics as a function of stellar mass, star formation rate, physical size, and light profile; the emergence of quiescent galaxies; the role of active nuclei; the impact of galaxies at the epoch of reionization; and the study of the earliest protogalaxies.

The full JADES observing programme [JADES Collaboration, in prep.] involves full use of coordinated parallel observations with both these instruments on JWST, resulting in extensive deep NIRCam data paired with both medium and deep NIRSpec multi-shutter spectroscopy, as well as some very deep MIRI pointings. Here we focus on the aspects relevant to the discoveries presented in this paper.

In GOODS-S, under GTO programme ID 1180 (PI: Eisenstein), JADES imaged the area immediately on and around the Hubble Ultra Deep Field, using 4 pointings to produce a deep, filled mosaic of roughly 4.4 by 6 arcminutes. In each pair of pointings, there is substantial overlap producing about 10 square arcmin of yet deeper imaging. In all 4 pointings, we conduct a 9-point dither of 1375 second individual exposures in each of 5 filter pairs. The filters are F090W, F115W (used twice), F150W, F200W in the short-wavelength (SW) channel, and F277W, F335M, F356W, F410M, and F444W in the long-wavelength (LW) channel. In 2 of the pointings, we also include an 8-point dither of 1375 s exposures in 3 filter pairs, adding depth to F090W, F115W, and F150W, F277W, F410M, and F444W. All imaging is at the same position angle of 298.5 degrees, in the period of Sept 29 to Oct 5, 2022. JADES is approved to double this depth, but only the first half was scheduled in year 1. The delay of the second half of the imaging program does mean that the SW chip gaps in the V2 'horizontal' direction are not yet covered at the time of this writing.

In the deepest regions, JADES year-1 imaging has reached 9.9, 16.8, 9.9, and 6.9 hrs of exposure time in F090W, F115W, F150W, and F200W, respectively, reaching 0.15" radius aperture point-source depths of approximately 4, 3.5, 3.8, and 4 nJy (5$\sigma$). These flux limits correspond to 29.89, 30.04, 29.95, and 29.89 AB magnitudes, respectively.

Most of these data were successful and met specifications. Two of the pointings suffer from unusually high persistence left from the previous program, which affects less than 10% of the pointing area and is largest in F090W. Mitigation of this problem is in progress, but the issues do not affect these results.

JADES is also using data from GO programme ID 1963 (PI Williams, Maseda & Tacchella) that provide additional imaging of a single NIRCam pointing that overlaps about 35% of the above mosaic. This program provides 7.7 hrs of integration in F182M, F210M, and F480M, as well as 3.9 hrs in F430M and F460M. This data was not used in the original selection of these candidates for NIRSpec MSA follow-up but has been co-reduced for their characterization.

JADES is observing this mosaic and other imaging with large amounts of NIRSpec spectroscopy. In particular, ref. 37 describe a single long NIRSpec observation, totaling 28 hours of integration. An additional 220 allocated hours of further GOODS-S spectroscopy is approved for later in JWST Cycle 1.

**Image Reduction**
The details of the data reduction and modelling, including performance tests, will be presented in ref. [ST in prep.]. We summarise here briefly the main steps. We process the raw images through the JWST Calibration Pipeline v1.8.1 with the CRDS pipeline mapping (pmap) context 1009, which includes in-flight NIRCam dark, distortion, bad pixel mask, read-noise, superbias reference, and ground flat (corrected for in-flight performance) files. Stage 1 of the JWST pipeline performs detector-level corrections and produces count-rate images. We run this step with the default parameters, including masking and correction of "snowballs" in the images caused by charge deposition following cosmic ray hits. Stage 2 of the JWST pipeline performs the flat-fielding and applies the flux calibration[52] (conversion from counts/s to Mjy/sr). We adopt the default values for these pipeline steps as well. Following this, we perform several custom corrections in order to account for several features in the NIRCam images[53]. Specifically, we measure and remove *1/f* noise, which is correlated noise introduced in the readout of the images[54]. The noise is visible as horizontal and vertical striping patterns. We measure the striping pattern using a sigma-clipped median along the rows and then the columns on the source-masked images. Afterwards, we subtract the "wisp" features from the short-wavelength channel images (in particular the NIRCam detectors A3, A4, B3 and B4 for the filters F115W, F150W, F182M, F200W, F210M). We have constructed wisp templates by stacking all images from our JADES (1180) programme and several other programmes (1063, 1345, 1837, 2738) after reducing them following the aforementioned approach. The templates are rescaled to account for the variable brightness of the wisp features and subtract them from the images. Following the wisp correction, we perform a background subtraction using the *photutils*[54] Background2D class. Before combining the individual exposures into a mosaic, we perform an astrometric alignment using a custom version of JWST TweakReg. We calculate both the relative and absolute astrometric correction for images grouped by visit and band by matching sources to a reference catalogue constructed from HST F160W mosaics in the GOODS-S field with astrometry tied to Gaia-EDR3[56]. We then run Stage 3 of the JWST pipeline, combining all exposures of a given filter and a given visit. We choose a pixel scale of 0.03 arcsec/pixel for both SW and LW channel images. We choose *drizzle*[57] parameters *pixfrac*=1 and 0.7 for the SW and LW images, respectively. These individual visit-level mosaics are then combined to create the final mosaic shown in Fig. 1.

**Detection and Photometry**
After the mosaics are created, a detection image is generated using *astropy*[58] by stacking with inverse variance-weighting the JWST FITS data model SCI flux extension and ERR flux error extension of the NIRCam F200W, F277W, F335M, F356W, F410M, and F444W filter images. The ERR extension is constructed from a quadrature sum of sky, read, and Poisson noise. The ratio of the weighted SCI and ERR stacks then serves as a measure of the local signal-to-noise ratio per pixel. We create a *photutils* catalogue by selecting contiguous regions of the detection image with SNR>3, and then apply a standard deblending algorithm[59] to the detection image with parameters *nlevels=32* and *contrast=0.001*. The structural properties of the sources are then computed from the SNR detection image, including location, size, ellipticity, on-sky orientation, and sizes[60]. A segmentation map is constructed to record regions of significant flux associated with each object.

Forced photometry is subsequently performed on the HST and JWST mosaics by measuring aperture fluxes at the locations of objects in the source catalogue without requiring a detection in each image and with a correction for aperture losses using model PSFs from the TinyTim[61] (HST) and WebbPSF[62] (JWST) packages assuming point source morphologies. Two error estimates for the flux are computed, including performing aperture photometry on the squared ERR extension to compute a flux variance and by using random apertures distributed across regions of the image. For the random aperture rms flux uncertainties, for each band 100,000 apertures distributed across the survey area are split into 100 groups of 1000 apertures based on the mean exposure time at the aperture locations. For each set of 1000 apertures, the rms flux in electrons among the apertures is measured as a function of aperture size. The relation between aperture size and rms flux is used to assign the contribution of the background sky to the flux uncertainty, and the background uncertainty is added in quadrature to the Poisson uncertainty to compute a total flux uncertainty for each object individually.

Table 2 presents the *forcepho* and aperture photometry for the four galaxies. The agreement between the photometric measurement methods is good overall, with typical agreement at the $1 - 2\sigma$ s.d. level. Characteristic fluxes in detected bands are several nJy for JADES-GS-z10-0, JADES-GS-z12-0, and JADES-GS-z13-0, while the brighter JADES-GS-z11-0 galaxy has fluxes $f_\nu \approx 10 - 20$ nJy. These objects are all strongly detected in F200W, with *forcepho* signal-to-noise ratios of $SNR_{F200W} \approx 20$. The strong detections in F200W enable a straightforward identification of the Lyman-$\alpha$ break photometrically, as discussed below.

**Photometric Redshift Analysis and Sample Selection**
To select these high-redshift galaxies, we perform a Lyman-break colour selection using the aperture photometry presented in Extended Data Table 1. Because of the opacity of the intergalactic medium to photons with rest wavelengths shorter than 1216 Ångstroms, to find galaxies at z > 9 we hunted for galaxies based on their colours, non-significant flux (less than $2\sigma$) in bands blueward of the proposed Lyman-$\alpha$ break, and detections with a $5\sigma$ significance in one or more filter images at longer wavelengths than the proposed Lyman-$\alpha$ break. Specifically, for F115W dropouts, we identified objects with F115W–F150W > 1.0 mag and F150W–F200W < 0.4 mag. For F150W dropouts, we searched for objects with F150W–F200W > 1.0 and F200W–F277W < 0.4 (see ref. 63). These colours permit the selection of galaxies with a Lyman−$\alpha$ break while also removing redder galaxies that may be quiescent or dusty lower-redshift interlopers.

To supplement the Lyman-break colour selection, we additionally calculated photometric redshifts using the template-fitting code EAZY[63] and the SED-fitting codes *Prospector*[42] and BEAGLE[65]. These fitting approaches use more photometric information than pure colour selection and allow for testing of the significance and reliability of the dropout selection. Fig. 3 shows the photometric redshift distributions inferred using Prospector, combining our NIRCam photometry in Extended Data Table 1. In each case, the spectroscopic redshifts confirmed by NIRSpec fall within the photometric redshift distributions. We also fit each of the candidate galaxies allowing redshift to be a free parameter and with a maximum redshift z< 6 to explore how the observed galaxy colours could be potentially replicated by a lower-redshift interloper. The new NIRCam sources presented here did not have probable interloper solutions at z<10 (see Fig. 3). We also generated stacks of the short-wavelength HST and JWST images for each candidate to further remove those objects with evidence of flux at wavelengths blueward of the proposed Lyman-break. Each high-redshift candidate was then inspected by multiple members of the collaboration independently, resulting in a catalogue of high-redshift candidate galaxies for NIRSpec follow-up.

**Detailed Model-fitting Photometry**
In addition to aperture photometry, we conduct model-fitting photometry on the selected sample of galaxies using the *forcepho* code [BDJ in prep.]. The *forcepho* code performs PSF-convolved scene modelling to the individual NIRCam exposures, providing a Markov Chain of simultaneously fit Sérsic profiles. In this way, it can provide a clear look at the flux correlations between multiple objects as well as avoid the image pixel correlations that result from a drizzled mosaic. Modelling all bands simultaneously leverages the higher spatial resolution of the shorter wavelength bands, while obtaining consistent colours that account for the varying PSF in each band. The samples of the posterior distribution for object fluxes yield uncertainty estimates that include the effects of neighbouring sources and uncertainty in the spatial profile.

For each object, we construct 2.5" by 2.5" cutouts from every overlapping NIRCam exposure after astrometric registration and photometric calibration are applied, but before any mosaicking or other pixel resampling. We mask pixels marked as "Do not use" by the reduction pipeline, including those identified as outliers during the mosaicking process. A sigma-clipped median of source-free regions is subtracted from each cutout. To account for any contamination by neighbouring sources, we simultaneously model all sources detected within these cutouts. The model for each source is a Sérsic profile[39] rotated, linearly stretched, and normalised by the total flux in each band. The 2D intrinsic profile of each source, which is forced to be the same in every band, is convolved with a Gaussian mixture approximation to the PSF in each NIRCam band to account for the different spatial resolutions. This model is then fit to all the imaging cutouts in every exposure at the same time, allowing the Sérsic profile for each source (with parameters half-light radius and Sérsic index) to vary in addition to the axis ratio, position angle, and total fluxes in each band. After the first round of optimization, the posterior distribution for all the parameters is sampled via Hamiltonian Monte Carlo[66]. Priors are uniform between 0.001" and 1" in half-light radius, between $0.8 < n < 6.2$ in Sérsic index, and between $0.4 < \sqrt{b/a} < 1.0$ in the square root of the axis ratio. Flux priors are uniform across a large range informed by the optimization results. Because of differences in the response between the long-wavelength detectors of NIRCam A and B modules we treat images through a single filter in these modules as separate bands. Samples of the posterior are used to estimate parameter values and their uncertainty. For the fluxes in Table 2 we report the mean and standard deviation of the samples, as the distribution is close to Gaussian.

In Extended Data Fig. 1 we show the posterior distributions for half-light radius and Sérsic index estimated from the MCMC samples. Each of the galaxies is well fit by a single Sersic profile. In all cases the bulk of the probability is at Sérsic index < 2. In two cases, JADES-GS-z13-0 and JADES-GS-z10-0 the radius distributions have substantial probability at the lower prior bound of 0.001", indicating that these sources are effectively unresolved. Fits with *forcepho* to unresolved brown dwarfs at similar F200W flux levels in our NIRCam data yield posterior distributions for size rising towards the lower bound we estimate 2-$\sigma$ s.d. for the half-light radius at 10 nJy of ~0.015". For the other two galaxies, JADES-GS-z11-0 and JADES-GS-z12-0, we infer sizes that are significantly above the lower bound (~0.02 and ~0.04" respectively, approximately 1 NIRCam pixel). We have also made fits to only the F200W data, which has the best combination of S/N and spatial resolution for these objects, and find consistent size constraints ruling out substantial biases from wavelength-dependent morphologies or astrometric offsets between bands.

**Stellar Population Synthesis Modelling**
Accounting for data quality cuts and spatial coverage, we retain fourteen bands of NIRCam photometry for JADES-GS-z11-0, eleven NIRCam bands for JADES-GS-z10-0 and JADES-GS-z13-0, and nine NIRCam bands for JADES-GS-z12-0. Together with the spectroscopic redshifts, this multiband photometry allows us the unprecedented opportunity to study the physical properties of a sample of galaxies only a few hundred Myr after the Big Bang. Nevertheless, the interpretation of galaxies' SEDs in terms of stellar population is complex, with many physical ingredients, and therefore we have utilised a number of modelling packages. These variations allow us to compare results and better appreciate the impact of assumptions and priors. The fiducial stellar population modelling in this work is done with the *Prospector* Bayesian SED fitting code[42]. We follow the procedures as outlined in ref. 67 and give a brief summary here. We then discuss our fiducial results and describe effects of prior assumptions and comparison to other SED fitting codes.

In our fiducial *Prospector* SED modelling, we assume the MIST stellar models[68] and a Chabrier initial mass function[69] (IMF) with a low- and high-mass cutoff of 0.08 $M_\odot$ and 120 $M_\odot$, respectively. We model the SED with a 13-parameter model that includes stellar mass (flat prior between $6 < log_{10} M_\star/M_\odot < 12$; defined as the integral of the SFH), stellar metallicity (a clipped normal prior in $log_{10} Z_\star/Z_\odot$ with mean $\mu = -1.5$, s.d. of $\sigma = 0.5$, with a minimum and maximum of -2.0 and 0.0, respectively) and gas-phase metallicity (a uniform prior between $-2.0 < log_{10} Z_\star/Z_\odot < 0.5$), a two-component dust attenuation model including birth-cloud dust attenuation and a diffuse component for the whole galaxy (3 parameters: diffuse dust optical depth describing the attenuation of all stellar light with a clipped normal prior with a mean $\mu = 0$, s.d. of $\sigma = 0.5$, min=0.0 and max=4.0; birth-cloud dust optical depth prior as a clipped normal relative to diffuse dust optical depth with a mean $\mu = 1$, s.d. of $\sigma = 0.3$, minimum of 0 and maximum of 2.0; and power-law modifier to shape of the Calzetti attenuation curve[70] of the diffuse dust with a uniform prior in the range of -1.0 and 0.4), and an ionisation parameter for the nebular emission (a uniform prior between $-4.0 < log_{10} U < 1.0$). The nebular emission is based on *cloudy*[71] photoionisation grids as described in ref. 72. In addition, we add a parameter ($f_{esc}$) that takes into account the absorption of H-ionising photons by dust or completely escaping the galaxy, for which we assume a clipped normal prior with a mean $\mu = 0$, a s.d. of $\sigma = 0.5$, a minimum of 0.0 and a maximum of 1.0. Furthermore, we adopt a flexible star-formation history (SFH) prescription[73] with 6 time bins. The SFH of the galaxies is assumed to start at redshift *z=20*. The first two time-bins are spaced at lookback times of 0-5 Myr and 5-10 Myr, while the other four are log-spaced out to $z = 20$. This flexible SFH prescription adds another 5 parameters. We assume the bursty-continuity prior[67] for the SFH, in which the prior for the logarithm of the ratio of the SFRs of two adjacent time-bins follows a student-t distribution with mean $\mu = 0$ and a s.d. of $\sigma = 1$. These statistics weakly imply a constant SFH prior. Given this physical model and its prior, we then sample the posterior distributions for all parameters with the dynamic nested sampling algorithm *dynesty*[74].

The motivation for those priors are the following. We assume a flexible SFH with a bursty prior because the SFHs of those early galaxies are expected to be bursty, i.e. vary on timescales of a few Myr[75,76]. We discuss assuming a parametric SFH and a less bursty, more continuous SFH prior below. We adopt an informative, low stellar metallicity prior but allow for up to solar metallicity since these galaxies could in principle enrich rapidly and it allows us to explore the degeneracy between *Z* and other parameters, including SFH and $f_{esc}$. Furthermore, we decouple the gas-phase metallicity from the stellar metallicity to account for possible alpha-element abundance enhancement[77] as well as out-of-equilibrium variations that are expected in bursty SFHs. We assume a flexible attenuation law to account properly for the uncertainties related in retrieving information (in particular about the SFH) from the rest-frame ultraviolet. Furthermore, the prior for attenuation is weighted towards low attenuation values as expected for early galaxies (consistent with the stellar metallicity prior). Finally the non-zero $f_{esc}$ parameter – accounting for both absorption and escape of H-ionising photons, but not for line ratio variations that can result from density-bounded HII regions – is motivated by the compactness of those galaxies[49,78,79].

Extended Data Fig. 2 shows the observed and derived posterior SEDs for our 4 *z*>10 galaxies. The blue circles show the detected photometric bands, while the arrows mark the upper limits. The red solid lines and shaded regions indicate the median and the 16-84[th] percentile of the posterior SED. We also show uncertainty normalised residuals ($\chi$) and note total $\chi^2$ for the most probable posterior sample. The observed photometric data are reproduced well overall by our inferred model. From these derived posterior SEDs, we calculate the UV continuum slope $\beta_{UV}$ for each object (Table 1). This calculation is done by taking the posterior SEDs, converting them to the rest frame, and fitting a slope to the spectrum in wavelength windows from ref . 80. These windows span 1268–2580 Å and are designed to omit spectral emission and absorption features. We perform this measurement of $\beta_{UV}$ for 1000 spectra drawn from the posteriors to include all model uncertainties. From these 1000 draws, we calculate the median value and the 68% credibility interval.

In Extended Data Figs. 3-6 we show, for each object, the joint posterior distributions for the stellar mass (integral of the SFH), specific SFR (sSFR) averaged over the past 30 Myr, stellar age $t_\star$, as defined by the lookback time when 50% of the stellar mass has formed (i.e. half-mass time), dust attenuation in the rest-frame V-band $A_V$, ratio of the dust attenuation in the rest-frame UV to the V-band (estimate of the attenuation law), escape fraction $f_{esc}$ (fraction of absorbed and escaped H-ionising photons), and stellar metallicity $Z_\star$. These figures highlight the uncertainties of individual key parameters and the degeneracies between them. For many parameters we obtain marginalised posterior distributions similar to the prior indicating the data are not constraining, in particular due to lack of information at wavelengths beyond the Balmer break. The half-mass ages are typically tens of Myr, but with highly non-Gaussian distributions often exhibiting a peak towards younger ages and a tail extending to 100 Myr. Degeneracies between age, dust attenuation, and metallicity are present. As expected, the inferences of stellar mass, half-mass age, stellar metallicity, and star formation are correlated with each other, with older ages tending to yield higher masses and lower sSFRs. In particular, lower metallicities ($log_{10} (Z_\star/Z_\odot) \sim$ -1.8) and stellar ages of $\sim$ 100 Myr as well as more moderate metallicities ($log_{10} (Z_\star/Z_\odot) \sim$ -1.3) and younger stellar ages can both fit the data. In the case of moderate metallicities, we infer a significant escape fraction of $f_{esc} \sim$ 0.5 to account for the non-detection of emission lines in both the photometry and spectroscopy (see also companion the paper ref. 37), which would be expected given those moderate metallicities. We find little attenuation for all galaxies ($A_V$ is consistent with 0.0-0.1 mag), but moderate attenuation values of up to 0.5 mag cannot be ruled out. Interestingly, we find that the posterior distribution of the attenuation law prefers rather shallow slopes with $A_{UV}/A_V \sim$ 1.5-2.5. The combination of all those effects lead to uncertainties on stellar mass of $\sim$0.3-0.7 dex.

In addition to this fiducial *Prospector* run, we run *Prospector* with two additional SFH priors: the standard continuity prior[72], which leads to more continuous and less bursty SFHs, and a parametric prior (delayed-tau model). The details of those two priors are described in ref. 67. Importantly, we assume the same priors for all other parameters. As expected[67,81], the standard continuity prior leads to slightly older ages (factors of 1.5-2.5) and hence higher stellar masses (by 0.0-0.2 dex). Other parameters do not change significantly. For the parametric delayed-tau prior, we find weakly increasing SFHs for all galaxies and the stellar ages and stellar masses agree all within a factor of 2 with our fiducial run. In summary, changing the SFH priors does not change our key results significantly: all changes are within the $1\sigma$ s.d. credibility intervals.

Finally, we also tested the robustness of the above results with two additional SED fitting codes: SEDz*[82] and BEAGLE[65,81]. Despite assuming different stellar libraries, isochrones and SFH modelling approaches, the stellar masses and the stellar ages from both SEDz* and BEAGLE agree with our fiducial ones when taking the credibility intervals into account. Specifically, SEDz* finds consistent stellar ages, but slightly higher stellar masses by about 0.3 dex. The stellar masses and ages of BEAGLE are consistent within the errors: For JADES-GS-z13-0, the BEAGLE stellar age is a factor of 3 older, leading to a stellar mass 0.15 dex lower (however, the

BEAGLE and Prospector values are still consistent with each other when considering the 1$\sigma$ s.d. credibility intervals). BEAGLE inferred a 1.5-times younger age and a 0.7 dex lower stellar mass than *Prospector* for JADES-GS-z12-0, but the older, more massive solution is also found in BEAGLE (and vice-versa). For the JADES-GS-z11-0, the ages are consistent, but the stellar masses are 0.6 dex lower in BEAGLE relative to *Prospector*. A reason for this difference could be that *Prospector* converges on a slightly decreasing SFH, while BEAGLE assumes a constant SFH. Finally, for JADES-GS-z10-0, BEAGLE inferred an age that is a factor of 3 older, while the stellar masses are consistent within 0.1 dex.

Extended Data Table 1| *forcepho* and Aperture Photometry

| NIRCam Band | JADES-GS+53.1499-27.7765 JADES-GS-z13-0 | | JADES-GS+53.1663-27.8216 JADES-GS-z12-0 | | JADES-GS+53.1648-27.7746 JADES-GS-z11-0 | | JADES-GS+53.1588-27.7735 JADES-GS-z10-0 | |
|---|---|---|---|---|---|---|---|---|
| | *forcepho* [nJy] | Aperture [nJy] | *forcepho* [nJy] | Aperture [nJy] | *forcepho* [nJy] | Aperture [nJy] | *forcepho* [nJy] | Aperture [nJy] |
| F090W | 0.36 ± 0.34 | -0.22± 0.87 | -2.01± 0.55 | 0.04± 0.97 | 0.20± 0.64 | -1.63± 2.03 | -0.13±0.35 | -0.24 ± 0.92 |
| F115W | 0.28 ± 0.25 | 0.59± 0.68 | -0.86 ±0.46 | -0.24 ±0.72 | -0.08 ±0.50 | -0.82 ±1.56 | -0.67±0.19 | -1.322 ±0.74 |
| F150W | -0.57± 0.27 | 0.18± 0.68 | 0.38± 0.42 | 0.42± 0.78 | 4.81± 0.67 | 5.87 ±2.48 | 5.95 ± 0.32 | 5.68 ± 0.78 |
| F182M | 5.11±0.69 | 7.69 ±1.55 | - | - | 14.57 ±0.72 | 15.81 ±1.86 | 7.83±0.33 | 9.83 ±0.94 |
| F200W | 6.43 ±0.38 | 6.13 ±0.80 | 8.50 ±0.58 | 9.14 ±0.80 | 14.93± 0.64 | 13.80 ±1.49 | 7.22 ± 0.31 | 7.11 ±0.68 |
| F210M | 6.87±0.75 | 8.57 ±1.95 | - | - | 15.08±0.80 | 13.04 ±2.09 | 7.60 ± 0.50 | 7.91 ± 1.05 |
| F277W | 6.35 ±0.50 | 6.09 ±0.41 | 8.83 ±0.45 | 9.02 ±0.47 | 17.52±0.74 | 17.58± 1.07 | 6.91 ± 0.39 | 6.91 ±0.61 |
| F355M | 4.91 ±0.90 | 4.89 ±0.76 | 5.26 ±0.67 | 7.40 ±0.93 | 11.56 ±1.17 | 13.26 ±1.80 | 5.49 ± 0.66 | 6.28 ± 0.81 |
| F356W | 5.44 ±0.56 | 4.42 ±0.49 | 8.47 ±0.50 | 8.69 ±0.57 | 15.55 ±0.75 | 16.26 ±1.06 | 5.36 ± 0.38 | 4.88 ±0.38 |
| F410M | 5.47 ±1.07 | 5.70 ±0.75 | 6.98 ±0.83 | 7.14 ±1.06 | 11.21 ±1.46 | 10.69 ±2.26 | 3.42 ±0.84 | 4.94 ±0.81 |
| F430M | - | - | - | - | 12.81±2.68 | 15.91± 3.00 | - | - |
| F444W | 5.68 ±0.84 | 4.79± 0.66 | 8.37 ±0.69 | 8.83± 0.82 | 15.98 ±1.212 | 13.75 ±1.73 | 7.40 ± 0.81 | 6.80 ± 0.72 |
| F460M | - | - | - | - | 13.04±3.94 | 14.06± 4.93 | - | - |
| F480M | - | - | - | - | 14.64±2.79 | 18.57 ±3.36 | - | - |

Columns: (1) NIRCam band, (2) *forcepho* photometry for JADES-GS-z13-0 in nJy, (3) *r=0.1"* radius circular aperture photometry for JADES-GS-z13-0 in nJy, (4) *forcepho* photometry for JADES-GS-z12-0 in nJy, (5) *r=0.15"* radius circular aperture photometry for JADES-GS-z12-0 in nJy, (6) *forcepho* photometry for JADES-GS-z11-0 in nJy, (7) *r=0.15"* radius circular aperture photometry for JADES-GS-z11-0 in nJy, (8) *forcepho* photometry for JADES-GS-z10-0 in nJy, (9) *r=0.1"* radius circular aperture photometry for JADES-GS-z10-0 in nJy.

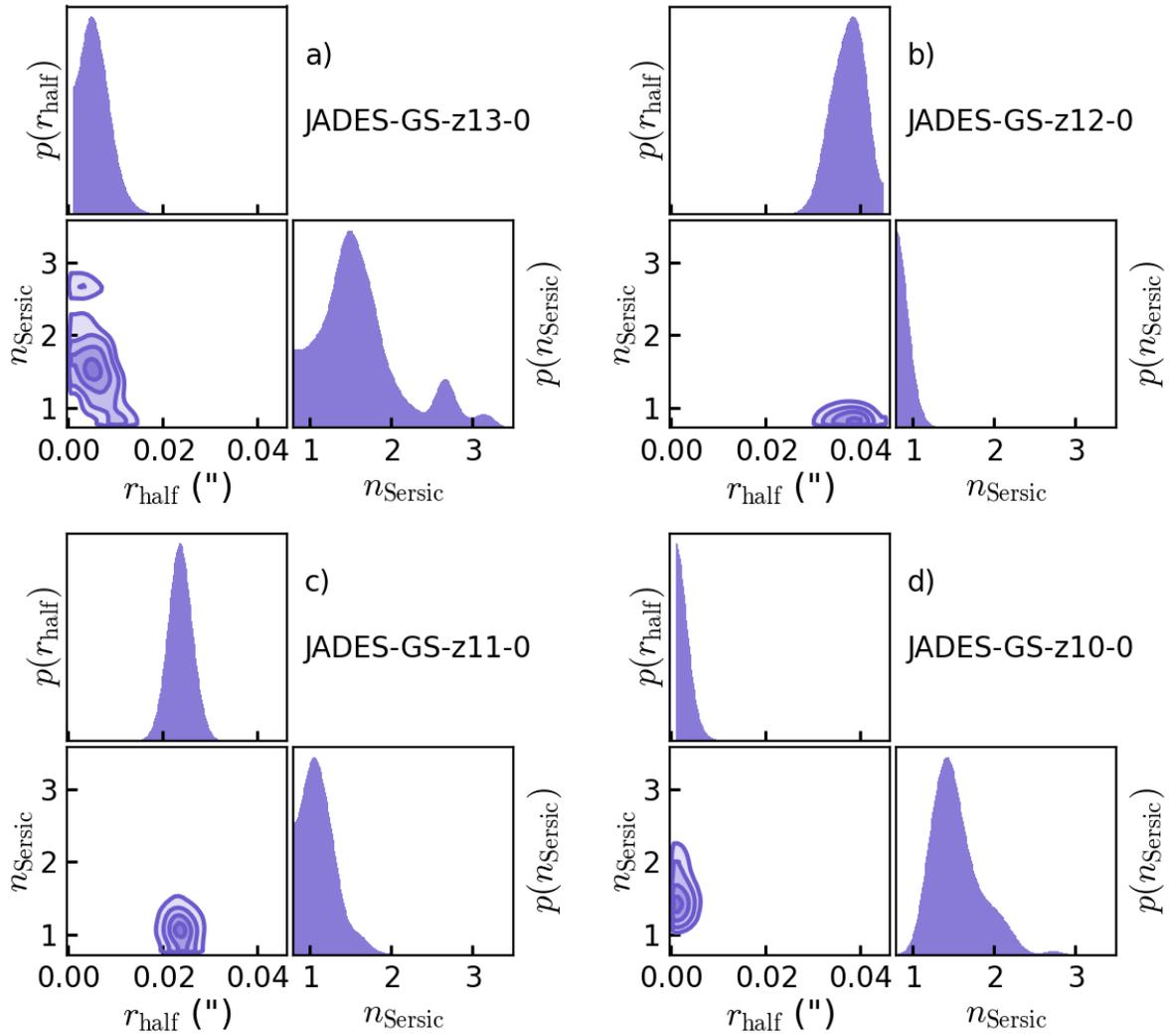

**Extended Data Fig. 1| Inferred spatial profiles.** The marginalised and joint posterior distribution for the Sérsic index and half-light radius from *forcepho* fitting to the individual exposures of each galaxy (**a-d**). We only infer an upper limit on the sizes of JADES-GS-z10-0 (**d**) and JADES-GS-z13-0 (**a**).

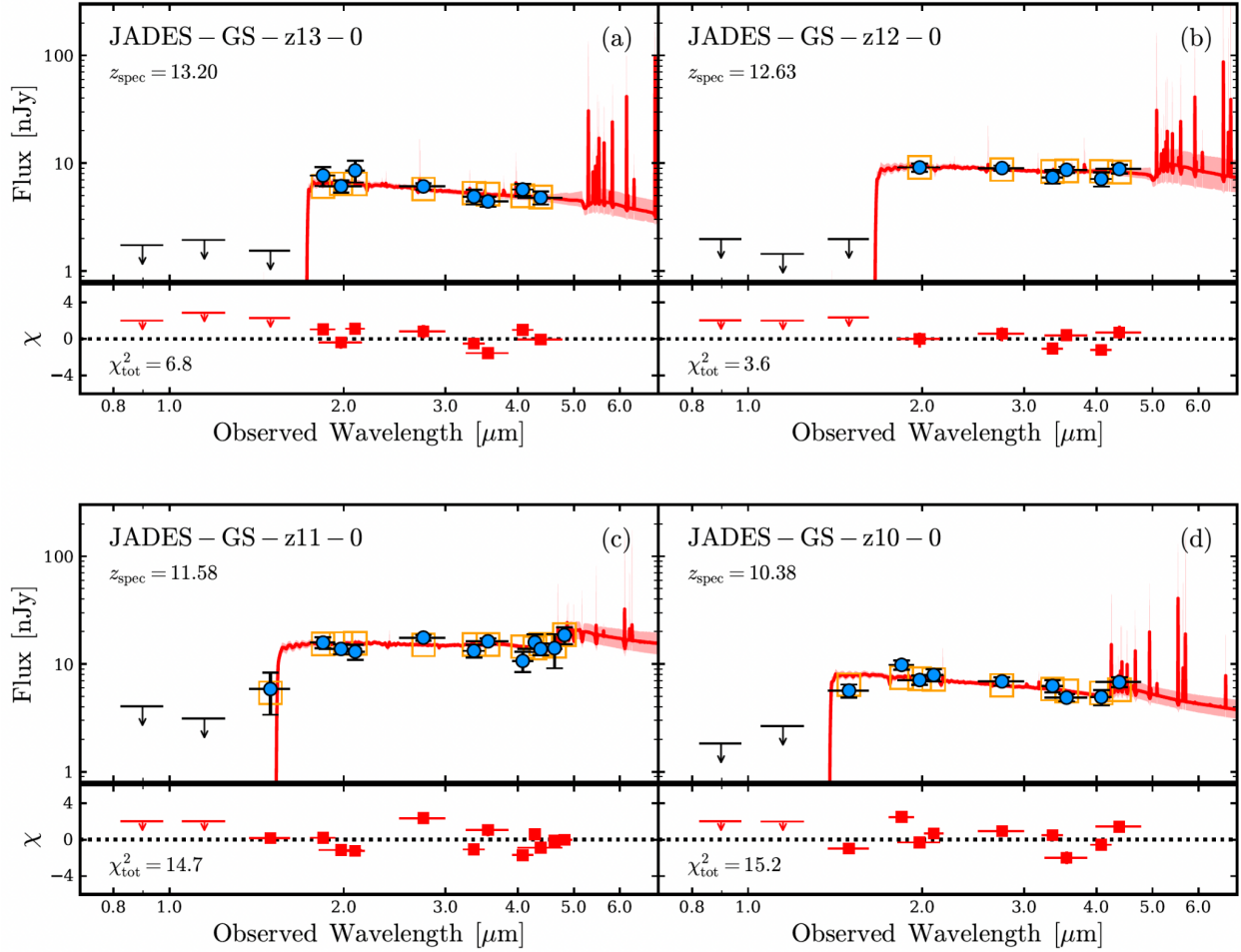

**Extended Data Fig. 2| Spectral energy distribution (SED) modelling of all four z>10 galaxies (panels a-d).** The observational data used in the SED fitting is shown in the top panels. The detected filters are plotted as blue circles, while the $5\sigma$ s.d. upper limits are indicated as bars with an arrow pointing down. Horizontal error bars indicate the approximate wavelength range of each filter. The red line with the shaded region around it shows the best-fitting SED with $1\sigma$ s.d. marginalised credibility interval. The orange boxes mark the posterior fluxes in the different filters. The bottom panels show the $\chi$ values for each filter and the total $\chi^2$ value is given.

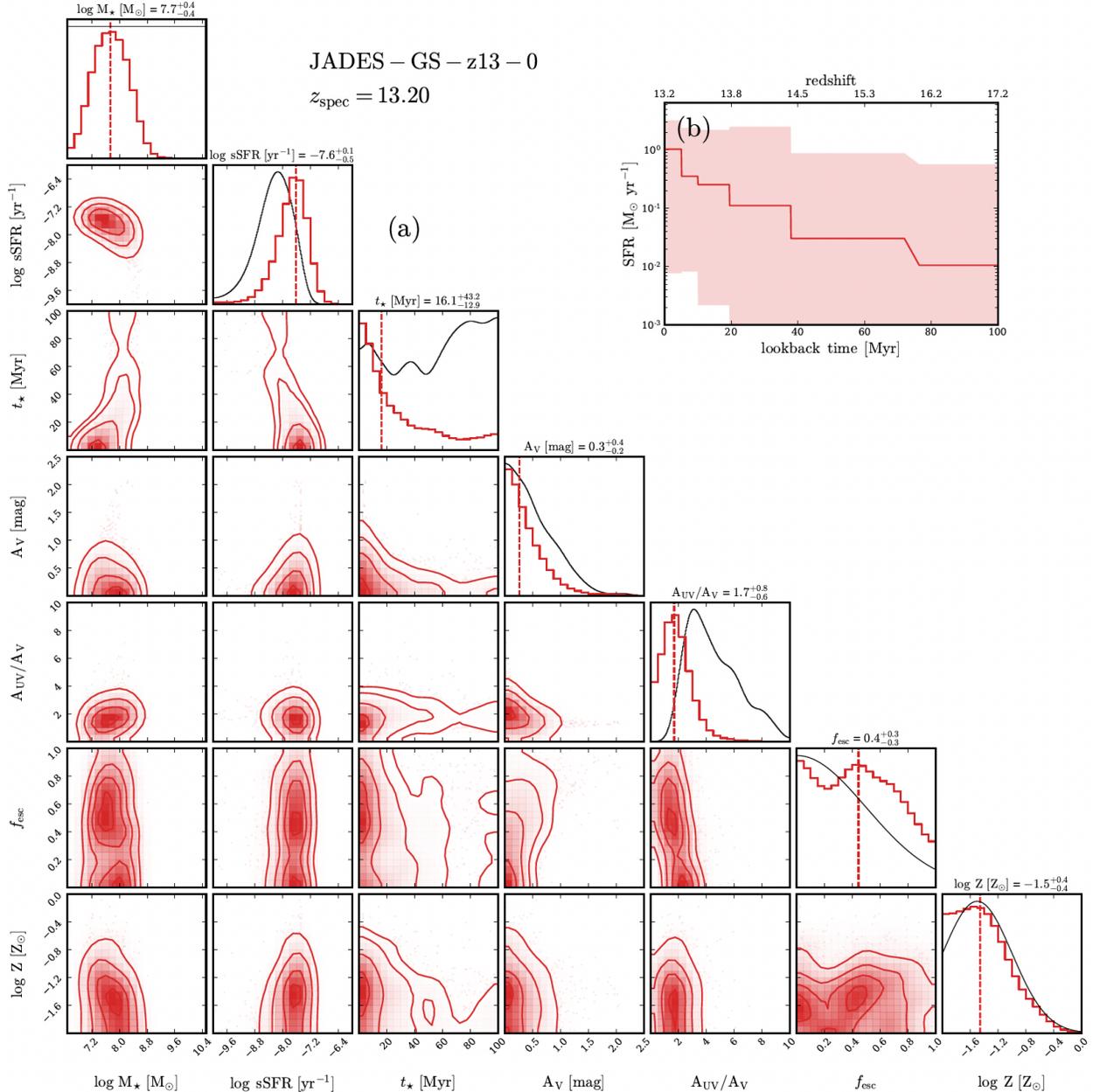

**Extended Data Fig. 3| Inferred posterior distributions for galaxy JADES-GS-z13-0 ($z_{\rm spec}$=13.20).** This corner figure (panel **a**) shows the posterior distribution of some of the key quantities that we infer from our SED modelling, including the stellar mass, specific SFR, stellar age (half-mass time), dust attenuation in the rest-frame V-band, the ratio of the attenuation in the rest-frame UV to the V-band (probing the slope of the attenuation law), escape fraction and stellar metallicity. The priors are indicated on the marginalised distributions as solid black lines. The inset on the top right (panel **b**) shows the posterior of the star-formation history (SFH). This galaxy is consistent with an increasing SFH and a young age, though older stellar populations of up to a 100 Myr cannot be ruled out. The SED of this galaxy is consistent with both a lower stellar metallicity solution (together with a low escape fraction and a steeper attenuation law) and higher metallicity solution (together with a higher escape fraction and a shallower attenuation law).

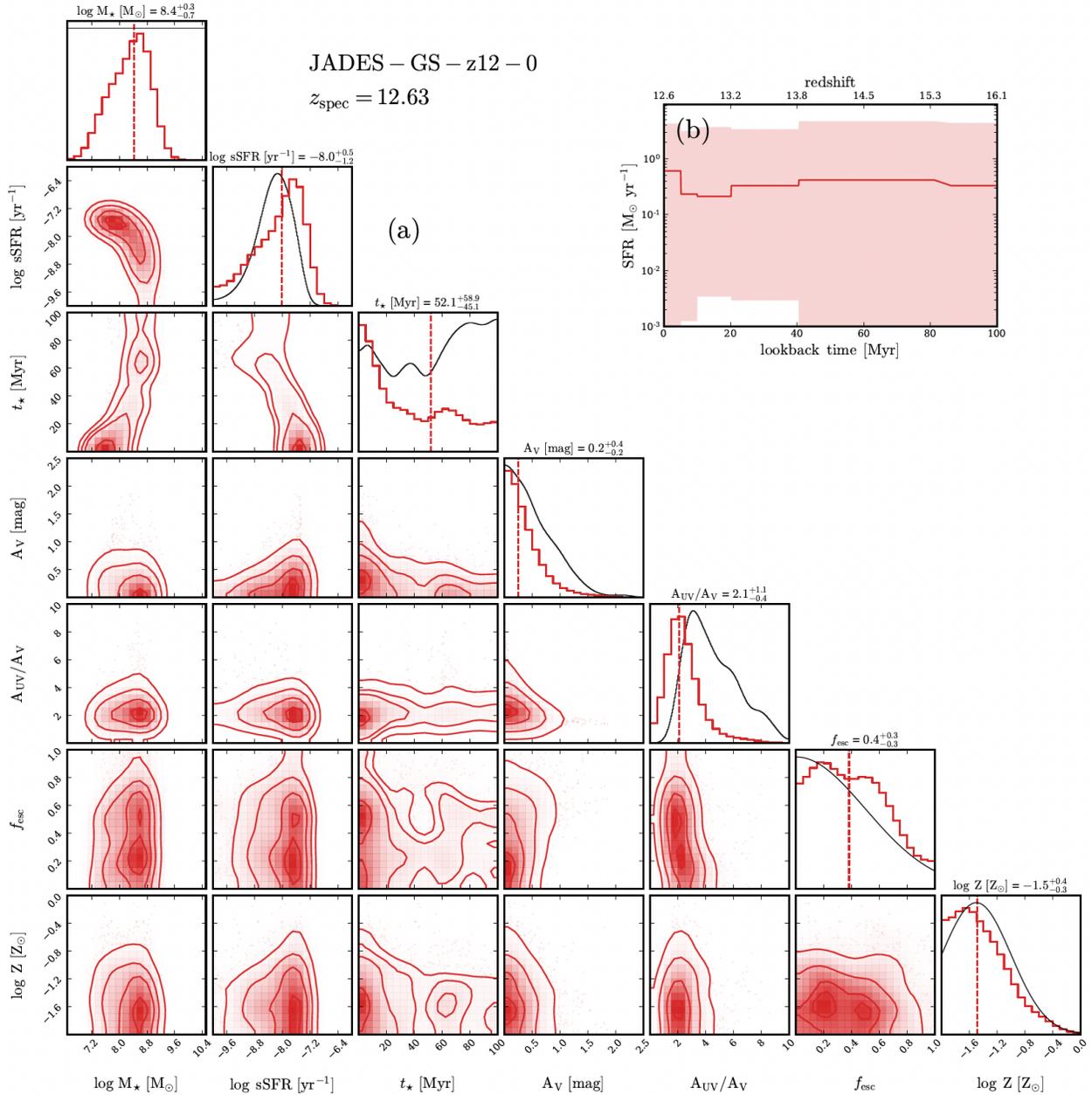

**Extended Data Fig. 4| Inferred posterior distributions for galaxy JADES-GS-z12-0 ($z_{spec}$=12.63).** The figure follows the same layout as Extended Data Fig. 3. The SFH of this galaxy is consistent with being constant. Although young ages are preferred, there is a significant tail to older ages, leading to a median of the posterior age distribution of 50 Myr.

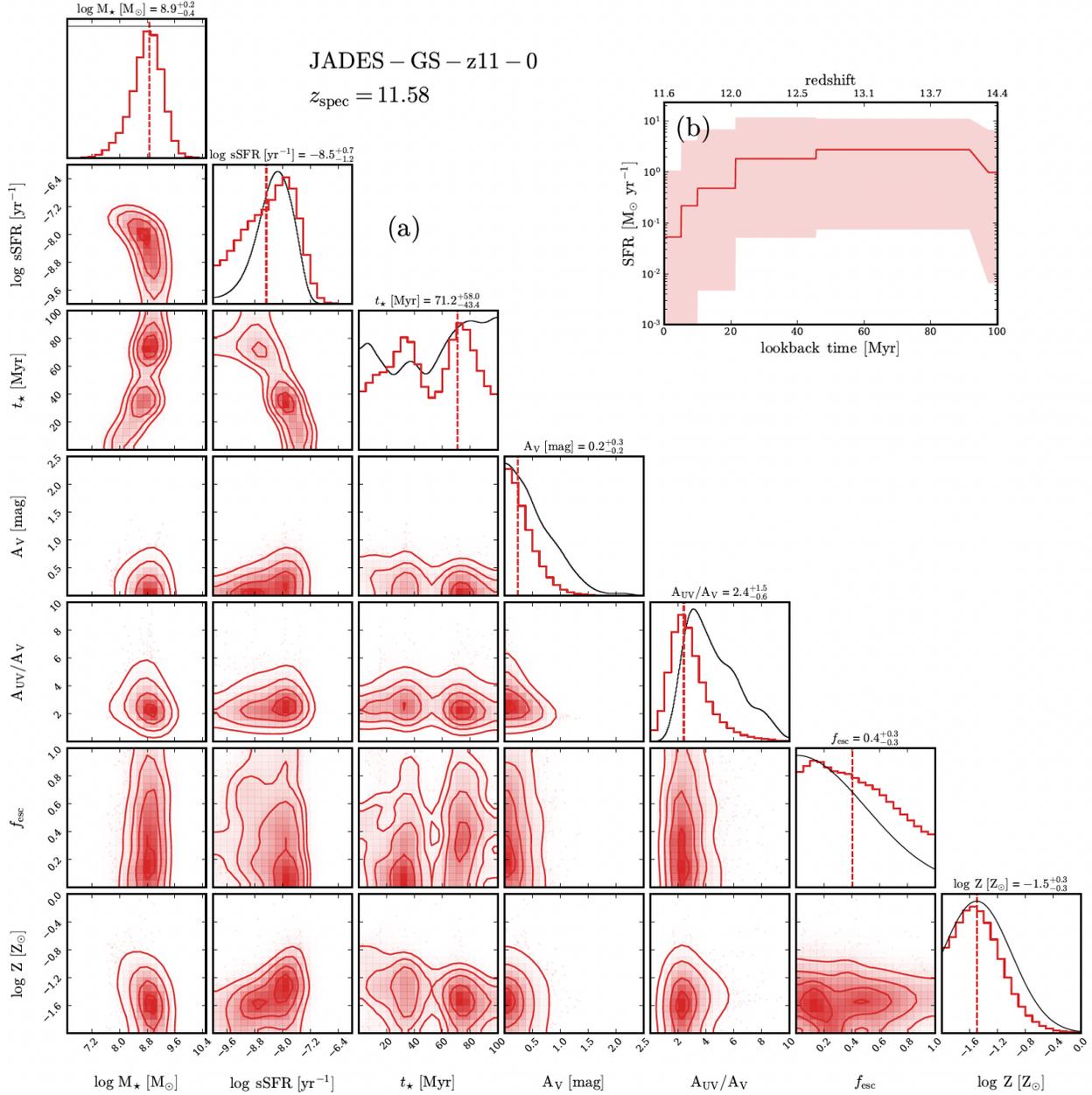

**Extended Data Fig. 5| Inferred posterior distributions for galaxy JADES-GS-z11-0 ($z_{spec}$=11.58).** The figure follows the same layout as Extended Data Fig. 3. The SED of this galaxy implies a significant older component with a decreasing SFH, leading to an age consistent with as high as 100 Myr.

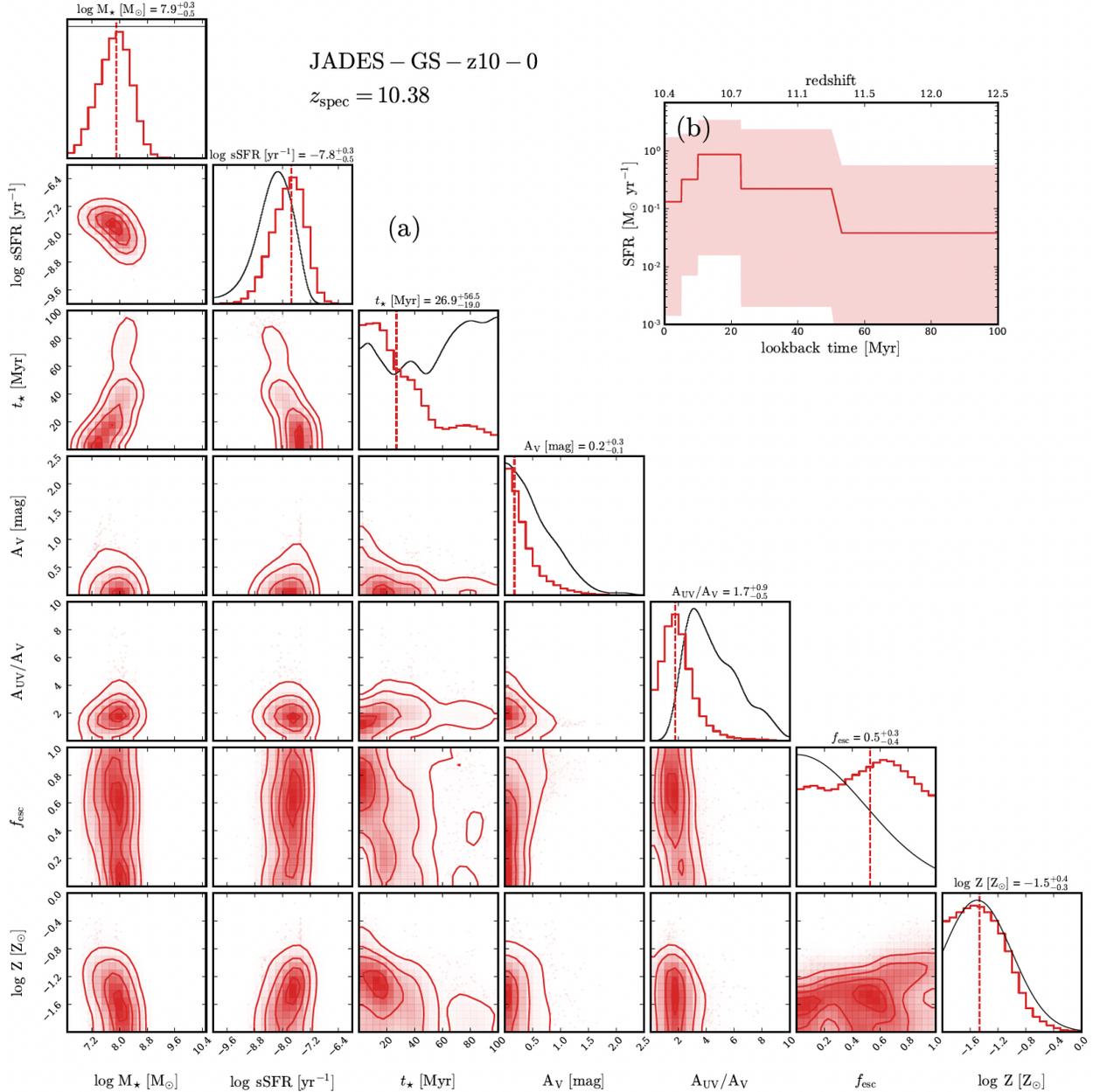

**Extended Data Fig. 6| Inferred posterior distributions for galaxy JADES-GS-z10-0 ($z_{spec}$=10.38).** The figure follows the same layout as Extended Data Fig. 3. The galaxy, similar to JADES-GS-z13-0, shows a bimodal distribution in the escape fraction.

**Data availability**
The data that support the findings of this study[83] are available from a public online repository.

**Code availability**
The *astropy*[58] and *photutils*[55] software suites are publicly available. The *Prospector* code[42] is publicly available, as is the *forcepho* photometric forward-modelling code.

**Acknowledgements**
BER, BDJ, DJE, MR, EE, GR, CNAW, MF, and FS acknowledge support from the JWST/NIRCam Science Team contract to the University of Arizona, NAS5-02015. LW acknowledges support from the National Science Foundation Graduate Research Fellowship under Grant No. DGE-2137419. DJE is further supported as a Simons Investigator. Sal acknowledge support from the James Webb Space Telescope (JWST) Mid-Infrared Instrument (MIRI) Science Team Lead, grant 80NSSC18K0555, from NASA Goddard Space Flight Center to the University of Arizona. RH was funded by the Johns Hopkins University, Institute for Data Intensive Engineering and Science (IDIES). GR acknowledges support from 80NSSC22K1293. Sar acknowledges support from the research project PID2021-127718NB-I00 of the Spanish Ministry of Science and Innovation/State Agency of Research (MICIN/AEI). NB and PJ acknowledge support from the Cosmic Dawn Center (DAWN), funded by the Danish National Research Foundation under grant no.140. AJB, AJC, JC, IEBW, & AS acknowledge funding from the "FirstGalaxies" Advanced Grant from the European Research Council (ERC) under the European Union's Horizon 2020 research and innovation programme (Grant agreement No. 789056). SC acknowledges support by European Union's HE ERC Starting Grant No. 101040227 – WINGS. ECL acknowledges support of an STFC Webb Fellowship (ST/W001438/1). MC, FDE, TJL, RM, JW, and LS acknowledge support by the Science and Technology Facilities Council (STFC), ERC Advanced Grant 695671 "QUENCH". RM is further supported by a research professorship from the Royal Society. JW is further supported by the Fondation MERAC. RS acknowledges support from a STFC Ernest Rutherford Fellowship (ST/S004831/1). HÜ gratefully acknowledges support by the Isaac Newton Trust and by the Kavli Foundation through a Newton-Kavli Junior Fellowship. SB acknowledges support from the Natural Sciences and Engineering Research Council (NSERC) of Canada. KB is supported in part by the Australian Research Council Centre of Excellence for All Sky Astrophysics in 3 Dimensions (ASTRO 3D), through project number CE170100013. REH acknowledges support from the National Science Foundation Graduate Research Fellowship Program under Grant No. DGE-1746060. This research made use of the *lux* supercomputer at UC Santa Cruz, funded by NSF MRI grant AST 1828315 (BER).

**Author contributions**
BER, ST, BDJ led the writing of this paper. MR, CNAW, EE, FS, GR, KH, CCW, and MF contributed to the design, construction, and commissioning of NIRCam. BER, ST, BDJ, CNAW, DJE, IS, MR, RE, Sal, ZC contributed to the JADES imaging data reduction. Rha, BER contributed to the JADES imaging data visualization. BDJ, ST, AD, DPS, LW, MWT, RE contributed the 24odelling of galaxy photometry. KH, JMH, JL, LW, RE, REH contributed the photometric redshift determination and target selection. BDJ, EN, KAS, ZC, ZJ contributed to the JADES imaging morphological analysis. BER, CNAW, CCW, KH, MR contributed to the

JADES pre-flight imaging data challenges. SCa, MC, JW, SAr contributed to the NIRSpec data reduction and to the development of the NIRSpec pipeline. PJ, NB, SAr contributed to the design and optimisation of the MSA configurations. AJC, AB, ECL, HU, KB, CNAW contributed to the selection, prioritisation and visual inspection of the targets. SCh, JC, ECL, RM, JW, RS, FDE, MM, MC, AdG, AS, LS contributed to analysis of the spectroscopic data, including redshift determination and spectral modelling. PJ, MS, TR, NL, NK contributed to the design, construction and commissioning of NIRSpec. FDE, TL, MM, MC, RM, SAr contributed to the development of the tools for the spectroscopic data analysis, visualisation and fitting. CW contributed to the design of the spectroscopic observations and MSA configurations. AB, AD, CNAW, CW, DJE, H-WR, MR, PF, PJ, RM, SAl, SAr contributed to the design of the JADES survey. SB commented on a draft of this paper. BER, CW, DJE, DPS, MR, NL, and RM serve as the JADES Steering Committee.

**Competing interests** The authors declare no competing interests.

**Additional Information**
**Correspondence and requests for materials** should be addressed to B. E. Robertson.
**Reprints and permissions information** is available at http://www.nature.com/reprints.

## References

1. Robertson, B. Galaxy Formation and Reionization: Key Unknowns and Expected Breakthroughs with James Webb Space Telescope. *Annu. Rev. Astron. Astrophys.* **60**, 121 (2022).

2. Carnall, A. C., et al. A first look at the SMACS0723 JWST ERO: spectroscopic redshifts, stellar masses and star-formation histories. Accepted to *Mon. Not. Royal Astron. Soc. Lett.* arXiv:220708778 (2022).

3. Brinchmann, J. High-z galaxies with JWST and local analogues – it is not only star formation. Submitted to *Mon. Not. Royal Astron. Soc.* arXiv:2208.07467 (2022).

4. Tacchella, S., et al. JWST NIRCam+NIRSpec: Interstellar medium and stellar populations of young galaxies with rising star formation and evolving gas reservoirs. Submitted to *Mon. Not. Royal Astron. Soc.* arXiv:2208.03281 (2022).

5. Curti, M., et al. The chemical enrichment in the early Universe as probed by JWST via direct metallicity measurements at $z \sim 8$. *Mon. Not. Royal Astron. Soc.* **stac2737** (2022).

6. Trussler, J. A. A., et al. Seeing sharper and deeper: JWST's first glimpse of the photometric and spectroscopic properties of galaxies in the epoch of reionisation. Submitted to *Mon. Not. Royal Astron. Soc.* arXiv:2207.14265 (2022).

7. Morishita, T. and Stiavelli, M. Physical characterization of F090W-dropout galaxies in the Webb's First Deep Field SMACS J0723.3-7323. Submitted to *Astrophys. J.*, arXiv:2207.11671 (2022).

8. Schaerer, D., et al. First Look with JWST spectroscopy: $z \sim 8$ galaxies resemble local analogues. *Astron. Astrophys. Lett.* **665**, L4 (2022).

9. Endsley, R., et al. A JWST/NIRCam Study of Key Contributors to Reionization: Star-forming and Ionizing Properties of UV-faint $z \sim 7-8$ Galaxies. Submitted to *Mon. Not. Royal Astron. Soc.*, arXiv:2208.14999 (2022).

10. Whitler, L., et al. On the ages of bright galaxies ~500 Myr after the Big Bang: insights into star formation activity at $z >\sim$ 15 with JWST. Submitted to *Mon. Not. Royal Astron. Soc.*, arXiv:2208.01599 (2022).

11. Topping, M., et al. Searching for Extremely Blue UV Continuum Slopes at $z = 7-11$ in JWST/NIRCam Imaging: Implications for Stellar Metallicity and Ionizing Photon Escape in Early Galaxies. Submitted to *Astrophys. J.*, arXiv:2208.01610 (2022).

12. Arellan-Cordova, K. Z. A First look at the Abundance Pattern – O/H, C/O, and Ne/O – in $z > 7$ Galaxies with JWST/NIRSpec. Accepted to *Astrophys. J.*, arXiv:2208.02562 (2022).

13. Inayoshi, K., et al. A Lower Bound of Star Formation Activity in Ultra-high-redshift Galaxies Detected with JWST: Implications for Stellar Populations and Radiation Sources. *Astrophys. J. Lett.* **938**, 10 (2022).

14. Ono, Y., et al. Morphologies of Galaxies at $z \simeq 9-17$ Uncovered by JWST/NIRCam Imaging: Cosmic Size Evolution and an Identification of an Extremely Compact Bright Galaxy at $z \sim 12$. Submitted to *Astrophys. J.* arXiv:2208.13582 (2022).

15. Rodighiero, G., et al. JWST unveils heavily obscured (active and passive) sources up to $z \sim$ 13. *Mon. Not. Royal. Astron. Soc. Lett.* arXiv:2208.02825 (2022).

16. Harikane, Y., et al. A Comprehensive Study on Galaxies at $z \sim 9-17$ Found in the Early JWST Data: UV Luminosity Functions and Cosmic Star-Formation History at the Pre-Reionization Epoch. Submitted to *Astrophys. J. Suppl. Ser.* arXiv:2208.01612 (2022).

17. Santini, P., et al. Early results from GLASS-JWST. XI: Stellar masses and mass-to-light ratio of $z > 7$ galaxies. Accepted to *Astrophys. J.* arXiv:2207.11379 (2022).

18. Leethochawalit, N., et al. Early results from GLASS-JWST. X: Rest-frame UV-optical properties of galaxies at $7 < z < 9$. Accepted by *Astrophys. J.* arXiv:2207.11135 (2022).

19. Roberts-Borsani, G., et al. A shot in the Dark (Ages): a faint galaxy at $z = 9.76$ confirmed with JWST. Submitted to *Nature*. arXiv:2210.15639 (2022).

20. Pontoppidan, K. M., et al. The JWST Early Release Observations. *Astrophys. J. Lett.* **936**, 14 (2022).

21. Finkelstein, S., et al. CEERS Key Paper I: An Early Look into the First 500 Myr of Galaxy Formation with JWST. Submitted to *Astrophys. J.* arXiv:2211.05792 (2022).

22. Treu, T., et al. The GLASS-JWST Early Release Science Program. I. Survey Design and Release Plans. *ApJ*, **935**, 110 (2022).

23. Castellano, M., et al. Early Results from GLASS-JWST. III. Galaxy Candidates at $z \sim 9-15$. *Astrophys. J. Lett.*, **938**, L15 (2022).

24. Adams, N. J., et al. Discovery and properties of ultra-high redshift galaxies ($9 < z < 12$) in the JWST ERO SMACS 0723 Field. Submitted to *Mon. Not. Royal Astron. Soc.*, arXiv:2207.11217 (2022).

25. Finkelstein, S. L., et al. A Long Time Ago in a Galaxy Far, Far Away: A Candidate $z \sim 12$ Galaxy in Early JWST CEERS Imaging, *Astrophys. J. Lett.* (in press), arXiv:2207.12474 (2022).

26. Naidu, R. P., et al. Two Remarkably Luminous Galaxy Candidates at $z \approx 10-12$ Revealed by JWST. *Astrophys. J. Lett.* (accepted)*,* arXiv:2207.09434 (2022).

27. Donnan, C. T., et al. The evolution of the galaxy UV luminosity function at redshifts $z \sim 8-15$ from deep JWST and ground-based near-infrared imaging. Submitted to *Mon. Not. Royal Astron. Soc.*, arXiv:2207.12356 (2022).

28. Hakim, A., et al. Revealing Galaxy Candidates out to $z \sim 16$ with JWST Observations of the Lensing Cluster SMACS0723. Submitted to *Mon. Not. Royal Astron. Soc.* arXiv:2207.12338 (2022).

29. Labbe, I., et al. A very early onset of massive galaxy formation. *Submitted to Nature.* arXiv:2207.12446 (2022).

30. Boylan-Kolchin, M., Stress Testing ΛCDM with High-redshift Galaxy Candidates. Submitted to *Mon. Not. Royal Soc. Lett.* arXiv2208.01611 (2022).

31. Lovell, C. C., et al. Extreme Value Statistics of the Halo and Stellar Mass Distributions at High Redshift: are JWST Results in Tension with ΛCDM? Accepted to *Mon. Not. Royal Soc.* arXiv:2208.10479 (2022).

32. Ferrara, A., Pallotini A., and Dayal, P. On the stunning abundance of super-early, massive galaxies revealed by JWST. arXiv:2208.00720 (2022).

33. Williams, H., et al. Spectroscopy from Lyman alpha to [O III] 5007 of a Triply Imaged Magnified Galaxy at Redshift $z = 9.5$. Submitted to *Astrophys. J.* arXiv:2210.15699 (2022).

34. Rieke, M., J., Kelly, D., Horner, S., Overview of James Webb Space Telescope and NIRCam's Role. SPIE **5904**, 1 (2005).

35. Jakobsen, P., et al. The Near-Infrared Spectrograph (NIRSpec) on the James Webb Space Telescope. I. Overview of the instrument and its capabilities. *Astronomy & Astrophysics*, **661**, 80 (2022)

36. Illingworth, G., et al. The Hubble Legacy Fields (HLF-GOODS-S) v1.5 Data Products: Combining 2442 Orbits of GOODS-S/CDF-S Region ACS and WFC3/IR Images, arXiv:1606.00841.

37. Curtis-Lake, E., Four metal-poor galaxies spectroscopically confirmed beyond redshift ten. *Submitted to Nature Astronomy*.

38. Bouwens, R. J., et al. A candidate redshift z~10 galaxy and rapid changes in that population at an age of 500 Myr. *Nature*. **469**, 504 (2011).

39. Ellis, R. S., et al. The Abundance of Star-forming Galaxies in the Redshift Range 8.5-12: New Results from the 2012 Hubble Ultra Deep Field. *Astrophys. J. Lett.* **763**, L7 (2013).

40. Bouwens, R. J., et al. Evolution of the UV LF from z~15 to z~8 Using New JWST NIRCam Medium-Band Observations over the HUDF/XDF. Submitted to *Mon. Not. Royal Astron. Soc.* arXiv:2211.02607 (2022).

41. Sérsic, J. L. Atlas de Galaxias Australes. Observatorio Astronomico, Cordoba, Argentina (1968).

42. Johnson, B., et al. Stellar Population Inference with Prospector. *Astrophys. J. Supp. Ser.* **254**, 22 (2021).

43. Wilkins, S. M., et al. First Light And Reionisation Epoch Simulations (FLARES) VII: The Star Formation and Metal Enrichment Histories of Galaxies in the early Universe. *Mon. Not. Royal Astron. Soc.* (submitted), arXiv2208.00976 (2022).

44. Tacchella, S., et al. A Redshift-independent Efficiency Model: Star Formation and Stellar Masses in Dark Matter Halos at z>4. *Astrophys. J.* **868**, 92 (2018).

45. Harris, J, Zaritsky, D. The Star Formation History of the Small Magellanic Cloud. *Astron. J.* **127**, 1531 (2004)

46. Weinberg, D. H., Andrews, B. H., Freudenburg, J. Equilibrium and Sudden Events in Chemical Evolution. *Astrophys. J.* **837**, 183 (2017)

47. Genzel, R., et al. A study of the gas-star formation relation over cosmic time. *Mon. Not. Royal Soc.* **407,** 2091 (2010)

48. Izotov, Y. I., et al. Low-redshift Lyman continuum leaking galaxies with high [O III]/[O II] ratios. *Mon. Not. Royal Soc.* **478** 4851 (2018)

49. Shapley, A. E., et al. Q1549-C25: A Clean Source of Lyman-Continuum Emission at z = 3.15. *Astrophys. J. Lett.* **826**, 24 (2016)

50. Sharma, M., et al. The brighter galaxies reionized the Universe. *Mon. Not. Royal Soc. L.* **458**, 94 (2016).

51. Giavalisco, M., et al. The Great Observatories Origins Deep Survey: Initial Results from Optical and Near-Infrared Imaging. *Astrophys. J.* **600,** L93-L98 (2004)

52. Boyer, M. L., et al. The JWST Resolved Stellar Populations Early Releases Science Program. I. NIRCam Flux Calibration. *Res. Not. Amer. Astron. Soc.* **6**, 191 (2022).

53. Rigby, J., et al. Characterization of JWST science performance from commissioning. arXiv:2207.05632 (2022).

54. Schlawin, E., et al. JWST Noise Floor. I. Random Error Sources in JWST NIRCam Time Series. *Astron. J.* **160**, 231 (2020).

55. Bradley, L., et al. astropy/photutils:1.5.0. Zenodo. https://doi.org/10.5281/zenodo.6825092 (2022)

56. Gaia Collaboration. Gaia Early Data Release 3. Summary of the contents and survey properties. *Astron. Astrophys.* **649,** 1 (2021).

57. Fruchter, A. S. and Hook, R. N. Drizzle: A Method for the Linear Reconstruction of Undersampled Images. *Pub. Astron. Soc. Pacif.*, **114**, 144-152 (2002).

58. Astropy Collaboration. The Astropy Project: Sustaining and Growing a Community-oriented Open-source Project and the Latest Major Release (v5.0) of the Core Package. *Astrophys. J.* **935,** 167 (2022).

59. Bertin, E. and Arnouts, S. SExtractor: Software for source extraction. *Astron. Astrophys. Supp.* **117**, 393-404 (1996).

60. Kron, R. G. Photometry of a complete sample of faint galaxies. *Astrophys. J. Supp. Ser.* **43**, 305-325 (1980).

61. Krist, J. E., Hook, R. N., Stoehr, F. 20 years of Hubble Space Telescope optical modeling using Tiny Tim. Proc. SPIE. 81270J (2011).

62. Perrin, M. D., et al., Updated point spread function simulations for JWST with WebbPSF, Proc. SPIE. 91433X (2014).

63. Hainline, K. N., et al. Simulating JWST/NIRCam Color Selection of High-redshift Galaxies. *Astrophys. J.* **892**, 125 (2020).

64. Brammer, G. B., van Dokkum, P. G., and Coppi, P. EAZY: A Fast, Public Photometric Redshift Code. *Astrophys. J.* **686**, 1503-1513 (2008).

65. Chevallard, J. and Charlot, S. Modelling and interpreting spectral energy distributions of galaxies with BEAGLE. *Mon. Not. Royal Astron. Soc.* **462**, 1415-1443 (2016).

66. Neal, R. MCMC Using Hamiltonian Dynamics. Handbook of Markov Chain Monte Carlo, Chapman & Hall/CRC, p. 113-162 (2011).

67. Tacchella, S., et al. On the Stellar Populations of Galaxies at $z\sim9-11$: The Growth of Metals and Stellar Mass at Early Times, *Astrophys. J.* **927**, 170 (2022).

68. Choi, J., Conroy, C., and Byler, N. The Evolution and Properties of Rotating Massive Star Populations. *Astrophys. J.* **838**, 159 (2017).

69. Chabrier, G. Galactic Stellar and Substellar Initial Mass Function. *Pub. Astron. Soc. Pacific.* **115**, 763-795 (2003).

70. Calzetti, D., et al. The Dust Content and Opacity of Actively Star-forming Galaxies. *Astrophys. J.* **533**, 682-695 (2000).

71. Ferland, G., et al. The 2013 Release of Cloudy. *Rev. Mex. Astron. Astrof.* **49**, 137-163 (2013).

72. Byler, N., et al. Nebular Continuum and Line Emission in Stellar Population Synthesis Models. *Astrophys. J.* **840**, 44 (2017).

73. Leja, J., et al. How to Measure Galaxy Star Formation Histories. II. Nonparametric Models. *Astrophys. J.* **876**, 3 (2019).

74. Speagle, J. S., DYNESTY: a dynamic nested sampling package for estimating Bayesian posteriors and evidences. *Mon. Not. Royal Astron. Soc.* **493**, 3132-3158 (2020).

75. Faucher-Giguère, C.-A., A model for the origin of bursty star formation in galaxies. *Mon. Not. Royal Astron. Soc.* **473**, 3717-3731 (2018).

76. Tacchella, S., Forbes, J. C., and Caplar, N., Stochastic modelling of star-formation histories II: star-formation variability from molecular clouds and gas inflow. *Mon. Not. Royal Astron. Soc.* **493**, 698-725 (2020).

77. Steidel, C. C., et al. Reconciling the Stellar and Nebular Spectra of High-redshift Galaxies. *Astrophys. J.* **826**, 159 (2016).

78. Ma, X., et al. No missing photons for reionization: moderate ionizing photon escape fractions from the FIRE-2 simulations. *Mon. Not. Royal Soc.* **498**, 2001-2017 (2020).

79. Naidu, R. P., et al. Rapid Reionization of the Oligarchs: The Case for Massive, UV-bright, Star-forming Galaxies with High Escape Fractions. *Astrophys. J.* **892,** 109 (2020).

80. Calzetti, D., Kinney, A. L., and Storchi-Bergmann, T. Dust Extinction of the Stellar Continua in Starburst Galaxies: The Ultraviolet and Optical Extinction Law. *Astrophys. J.* **429**, 582-601 (1994).

81. Whitler, L., et al. Star formation histories of UV-luminous galaxies at z~6.8: implications for stellar mass assembly at early cosmic times. *Mon. Not. Royal Astron. Soc.* **519,** 151-171 (2023).

82. Dressler, A., et al. Early Results from GLASS-JWST. XVII: Building the First Galaxies – Chapter 1. Star Formation Histories at 5 < *z* < 7. Submitted to *Astrophys. J. Lett.*, arXiv:2208.04292 (2022).

83. Robertson, B. E., Tacchella, S., Johnson, B. D., & JADES Collaboration. (2022). JWST Advanced Deep Extragalactic Survey: Thumbnail Images of Spectroscopically-Confirmed Galaxies at z>10. (v0.5.1) [Data set]. Zenodo. https://doi.org/10.5281/zenodo.7460116